\documentclass[sigconf]{acmart}
\usepackage{graphicx}
\usepackage{graphicx}
\usepackage[show]{chato-notes}
\usepackage{amsmath}
\usepackage{epsfig}
\usepackage{algorithm}
\usepackage[noend]{algorithmic}
\usepackage{url}
\usepackage{hyperref}
\usepackage{boldline}
\usepackage{xspace}
\usepackage{tikz}
\usetikzlibrary{shadows}
\usepackage{xcolor}
\usepackage[framemethod=TikZ]{mdframed}
\usepackage{float}
\usepackage{wrapfig}
\usepackage{xcolor}
\usepackage{adjustbox}

\newcommand{\spara}[1]{\smallskip\noindent{\bf #1}}

\newtheorem{problem}{Problem}

\newcommand{\obs}{\textsf{OBS}}
\newcommand{\dbs}{\textsf{DBS}}
\newcommand{\ds}{\textsf{DS}}
\newcommand{\ofs}{\textsf{OFS}}
\newcommand{\dfs}{\textsf{DFS}}

\DeclareMathOperator*{\argmin}{arg\,min}

\newcommand{\td}{\textsf{TD}\xspace}
\newcommand{\asd}{\textsf{ASD}\xspace}

\newcommand{\roi}{\ensuremath{\text{ROI}}\xspace}
\newcommand{\mri}{\ensuremath{\text{MRI}}\xspace}

\newcommand{\squishlist}{
 \begin{list}{$\bullet$}
  {  \setlength{\itemsep}{0pt}
     \setlength{\parsep}{3pt}
     \setlength{\topsep}{3pt}
     \setlength{\partopsep}{0pt}
     \setlength{\leftmargin}{2em}
     \setlength{\labelwidth}{1.5em}
     \setlength{\labelsep}{0.5em}
} }
\newcommand{\squishlisttight}{
 \begin{list}{$\bullet$}
  { \setlength{\itemsep}{0pt}
    \setlength{\parsep}{0pt}
    \setlength{\topsep}{0pt}
    \setlength{\partopsep}{0pt}
    \setlength{\leftmargin}{2em}
    \setlength{\labelwidth}{1.5em}
    \setlength{\labelsep}{0.5em}
} }

\newcommand{\squishdesc}{
 \begin{list}{}
  {  \setlength{\itemsep}{0pt}
     \setlength{\parsep}{3pt}
     \setlength{\topsep}{3pt}
     \setlength{\partopsep}{0pt}
     \setlength{\leftmargin}{1em}
     \setlength{\labelwidth}{1.5em}
     \setlength{\labelsep}{0.5em}
} }

\newcommand{\squishend}{
  \end{list}
}

\copyrightyear{2021}
\acmYear{2021}
\setcopyright{acmlicensed}\acmConference[KDD '21]{Proceedings of the 27th ACM SIGKDD Conference on Knowledge Discovery and Data Mining}{August 14--18, 2021}{Virtual Event, Singapore}
\acmBooktitle{Proceedings of the 27th ACM SIGKDD Conference on Knowledge Discovery and Data Mining (KDD '21), August 14--18, 2021, Virtual Event, Singapore}
\acmPrice{15.00}
\acmDOI{10.1145/3447548.3467154}
\acmISBN{978-1-4503-8332-5/21/08}
\begin{document}
\fancyhead{}
\title[Counterfactual Graphs for Explainable Classification of Brain Networks]{Counterfactual Graphs for Explainable Classification\\ of Brain Networks }

\author{Carlo Abrate}
\affiliation{%
	\institution{Sapienza University, Rome, Italy}
\institution{ISI Foundation, Turin, Italy}
}
\email{carlo.abrate@isi.it}

\author{Francesco Bonchi}
\affiliation{%
	\institution{ISI Foundation, Turin, Italy}
\institution{Eurecat, Barcelona, Spain}
}
\email{francesco.bonchi@isi.it}

\begin{abstract}
Training graph classifiers able to distinguish between healthy brains and dysfunctional ones,
can help identifying substructures associated to specific cognitive phenotypes. However, the mere predictive power of the graph classifier is of limited interest to the neuroscientists, which have plenty of tools for the diagnosis of specific mental disorders. What matters is the interpretation of the model, as it can provide novel insights and new hypotheses.

In this paper we propose \emph{counterfactual graphs} as a way to produce local post-hoc explanations of any black-box graph classifier. Given a graph and a black-box, a counterfactual is a graph which, while having high structural similarity with the original graph, is classified by the black-box in a different class. We propose and empirically compare several strategies for counterfactual graph search. Our experiments against a white-box classifier with known optimal counterfactual, show that our methods, although heuristic, can produce counterfactuals very close to the optimal one. Finally, we show how to use counterfactual graphs to build global explanations correctly capturing the behaviour of different black-box classifiers and providing interesting insights for the neuroscientists.
\end{abstract}


\maketitle \sloppy

\section{Introduction}
\label{sec:intro}
In recent years \emph{brain connectomics}~\cite{sporns2005human}, a field concerned with providing a network representation of the brain by comprehensively mapping the neural elements and their connections, has emerged as an important paradigm promising to help understanding mental processes and brain diseases \cite{FORNITO2015}. Thanks to the network representation of the brain, many interesting neuroscience challenges can be tackled as graph mining problems~\cite{bullmore2009complex}.
A prominent example is that of \emph{classification of brain networks}: given two groups of individuals,
a \emph{condition group}  (class 1)  and a \emph{control group} (class 0), where each individual is represented as a brain network, i.e., a graph defined over the same set $V$ of vertices corresponding to the brain \emph{regions of interest} ({\roi}s), the goal is to infer a model that, given an unseen graph defined over the same vertices $V$, it is able to predict whether the new individual belongs to class 0 or class 1 ~\cite{wang2017brainclassification,meng2018brainclassification,yan2019groupinn,lanciano2020}.

In this setting, building accurate classifiers is important, but not the main goal.
In fact, the mere predictive power is of limited interest to the neuroscientists, which have plenty of tools for the diagnosis of specific mental disorders. What matters is the ability of discovering discriminative substructures that might be associated to specific cognitive phenotypes
or mental dysfunctions~\cite{du2018brainclassificationsurvey}, so to provide insights and new hypotheses for the neuroscientists to further investigate.
Surprisingly, not much attention has been devoted in the literature to produce explanations for graph classification.

On the one hand, most of the methods proposed for graph classification are not very transparent and enjoy little interpretability, as they are based either on kernels \cite{shervashidze2011weisfeiler}, embeddings \cite{annamalai2017graph2vec,adhikari2018sub2vec, gutierrez2019embedding}, or deep learning \cite{yanardag2015deepWL, lee2018graphclass, wu2019demonet}.
On the other hand, although providing \emph{post-hoc explanations of a black-box classifier} is  a very active research area~\cite{guidotti2018survey,gilpin2018explaining}, most of the proposed methods focus on tabular data~\cite{ribeiro2016should,lundberg2017unified}, images~\cite{selvaraju2017grad,shrikumar2017learning,jimaging6060052} or time series~\cite{assaf2019explainable}.

\begin{figure}[t]
\vspace{-1mm}
\begin{minipage}{0.15\textwidth}
  \includegraphics[width=1.3\linewidth]{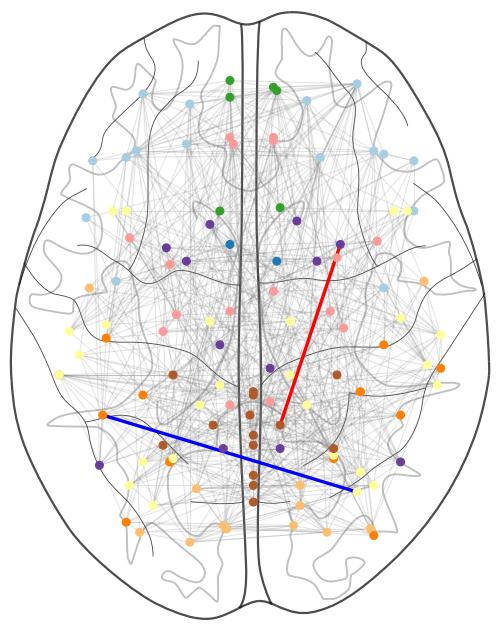}
\end{minipage}
\begin{minipage}{0.32\textwidth}
\centering
\includegraphics[width=.64\linewidth]{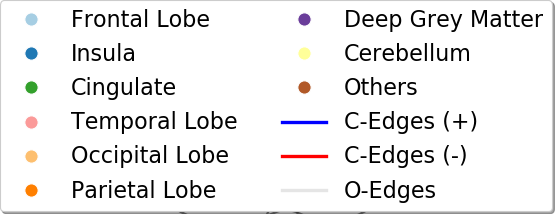}

\medskip \medskip \medskip
\medskip \medskip \medskip

\begin{mdframed}[innerbottommargin=3pt,innertopmargin=3pt,innerleftmargin=1pt,innerrightmargin=1pt,backgroundcolor=gray!10,roundcorner=10pt]
\begin{scriptsize}
\begin{center}
\emph{Patient \textsf{USM\_0050453} is classified as \asd (class 1). \\
If the connection between \textbf{Putamen\_R} and \textbf{Cerebelum\_9\_R} did not exist and instead the connection between \textbf{Parietal\_Inf\_L}  and \textbf{Cerebelum\_Crus2\_R} existed, then the patient would have been classified as \td (class 0).}
\end{center}
\end{scriptsize}
\end{mdframed}
\end{minipage}

\vspace{-5mm}

\caption{An example of counterfactual graph for a patient from the ABIDE dataset \cite{craddock2013abide}.
The patient is classified by a black-box (see Appendix \ref{sec:appendix}) in class \asd (Autism Spectrum Disorder), but with just two edge changes could have been classified as \td (Typically Developed). In gray the edges of the original graph, in red the edge to be removed and in blue the one to be added, so to form the counterfactual graph. In the box, the counterfactual expressed in natural language.}
    \label{fig:intro}
    \vspace{-1mm}
\end{figure}

In this paper, we propose \emph{graph counterfactuals} as a way to produce local post-hoc explanations of any black-box graph classifier. Counterfactuals are borrowed from causal language to create contrastive example-based explanations, similar to the following~\cite{counterfactuals}:

\begin{center}
\emph{``You were denied a loan because your annual income was \pounds 30,000. If your income had been \pounds 45,000, you would have been offered a loan."}
\end{center}

Intuitively, a counterfactual for a given data point $x$ is another data point $x'$ which is as close as possible to $x$, yet classified in a different class. The rationale for finding a counterfactual instance which is as close as possible to the original example is twofold: firstly, it ensures that the counterfactual explanation is compact, thus much easier to communicate and comprehend; secondly, it guarantees that the produced counterfactual is a realistic example, being not too far from a real data point.

In the context of brain network classification, we create a counterfactual graph, by changing the links (either adding or removing) of a given graph.  Figure \ref{fig:intro} provides an example of such counterfactual graph. Patient \textsf{USM\_0050453} from the ABIDE dataset \cite{craddock2013abide} (more information in Section \ref{sec:experiments} and Appendix \ref{sec:appendix}) is predicted in class \asd (Autism Spectrum Disorder, class 1) by a black-box graph classifier. By just removing an edge and adding another non-existing edge, we can obtain a new graph, which is very similar to the original one, but classified differently by the same black-box classifier.

\spara{Paper contributions and roadmap.} Besides the importance of explainability, classification of brain networks distinguishes from other graph classification tasks for other aspects. As discussed by other authors, e.g.,  \cite{gutierrez2019embedding,lanciano2020}, brain networks enjoy \emph{node identity awareness}, i.e.,  the fact that a specific vertex id is unique in a graph and corresponds to the same brain \roi through all the input networks, which are all thus defined  over the same vertices $V$. This is different from other graph classification tasks and other application domains (e.g., prediction of functional and structural properties of molecules in cheminformatics) where vertices can share the same labels and the vertex identity is not a relevant information.

In this specific context, the roadmap and contributions of this paper are as follows:
\squishlist
\item We define the \textsc{Counterfactual Graph Search} problem (\S\ref{sec:preliminaries}). As the search space of the problem is exponential and we do not know anything about the black-box classifier, we can not hope for a polynomial-time exact algorithm.
\item We thus propose heuristic-search methods  (\S\ref{sec:algorithms}), distinguishing between two cases: when we do not have any data (oblivious), or when we have access to a dataset (data-driven).
\item Our empirical comparison on different brain network datasets and with different black-boxes (\S\ref{subsec:exp:comparison}) show that both the oblivious and the data-driven approaches can produce good quality counterfactual graphs, but that the data-driven approach can do so faster (i.e., with less calls to the black-box classifier).
\item Our experiments against a \emph{white-box classifier with known optimal counterfactual}  (\S\ref{subsec:exp:whitebox})   show that our heuristic methods can produce counterfactuals very close to the optimal one.
\item We finally show how counterfactual graphs can be used as basic building blocks to produce  LIME-like~\cite{ribeiro2016should} (SHAP-like~\cite{lundberg2017unified}) local explanations  (\S\ref{subsec:appl:local}), as well as global explanations (\S\ref{subsec:appl:global1} and \S\ref{subsec:appl:global2}) correctly capturing the behaviour of different black-box classifiers and which are meaningful from the neuroscience standpoint.
\squishend

\section{Related work}
\label{sec:related}
As discussed in Section \ref{sec:intro} not much effort has been devoted to explainability in graph classification. Explanation methods for node classification and link prediction have been proposed in  \cite{ying2019gnnexplainer,zhang2020relex,huang2020graphlime}, but this is a different problem from
the graph classification task in which the aim is to classify the whole graph and not its nodes.

In the context of brain networks, Yan et al.
\cite{yan2019groupinn} propose a deep learning approach for explainable classification, using a node-grouping layer before the convolutional layer:  such node-grouping layer provides insight in the most predictive brain subnetworks.

Still in the context of brain network classification, Lanciano et al.~\cite{lanciano2020} propose a method based on \emph{contrast subgraph}, i.e., a set of vertices whose induced subgraph is very dense in one class of graphs and very sparse in the other class.
This approach produces very simple classifiers which are transparent and self-explanatory.

While \cite{yan2019groupinn,lanciano2020} propose methods that have some explainability \emph{by-design}, here we study how to produce local \emph{post-hoc} explanations of \emph{any} black-box graph classifier, by means of graph counterfactuals.

In \emph{cheminformatics}, Numeroso and Bacciu \cite{molecularcounterfactuals}  propose molecule counterfactuals for explaining \emph{deep graph networks} for the prediction of functional and structural
properties of molecules. Their proposal is an explanatory agent based on \emph{reinforcement learning} which has access to the internal representation of the black-box and leverages domain knowledge to constrain the generated explanations to be valid molecules.
The main difference with our work is in the type of graph classification assumed: our graphs, in fact, enjoy \emph{node identity awareness} \cite{gutierrez2019embedding,lanciano2020}, i.e.,  the fact that a specific vertex id is unique in a graph and corresponds to the same brain \roi through all the input networks. Conversely, molecules are small graphs where the vertices are atoms, the same vertex can thus appear many times in the same graph, and the links are labeled with the type of bond. Moreover, our approach provides explanations for any black-box classifier, not only for a specific one; it does not assume any knowledge of the internals of the black-box, nor it needs any background knowledge by the domain expert.

\section{Preliminaries}
\label{sec:preliminaries}
Magnetic resonance imaging ({\mri}) has played a central role in the development of connectomics,
 by allowing relatively cost-effective in vivo assessment of the macro-scale architecture of brain network connectivity. A connectome, or brain network, is created by linking the brain regions of interest ({\roi}s),
either by observing anatomic fiber density (\emph{structural connectome}), or
by computing pairwise correlations between time series of activity associated to {\roi}s (\emph{functional connectome}). The latter approach, known as f{\mri},
exploits the link between neural activity and blood flow and oxygenation,
to associate a time series to each {\roi}. A brain network can then be defined by creating a link between two {\roi}s that exhibit co-activation, i.e., strong correlation in their time series.

{\roi}s are defined by aggregating adjacent \emph{voxels}, i.e., the most fine-grain units of the 3-dimensional image which correspond to a very large number of brain cells (in the order of millions). This is done as a dimensionality reduction step and it is justified by the fact that spatially adjacent voxels typically play their role on the same activities, acting as a unique unit.
For this voxel-aggregation task, named \emph{parcellation} of the brain, several atlases are well established in neuroscience \cite{aal,craddock2012threshold}.
Finally, as \mri signals are heavily subject to noise caused by different confounding factors, a plethora of signal pre-processing strategies (e.g., filtering, signal correction, thresholding etc.)
are typically adopted in the data preparation pipeline (see Lang et al.~\cite{lang2012preprocessing}
for a comprehensive survey).

For the purposes of this paper, we abstract from all the pre-processing details (except when providing the experimental setting information needed for repeatability), and just assume that each brain is given to us as an undirected unweighted simple graph. More formally we consider a set of vertices  $V$, i.e., the  {\roi}s, as defined by the selected parcellation atlas. We let  $V^2$ denote the set of all possible pairs of elements of $V$, i.e., $V^2 = \{(u,v) | u, v \in V \wedge u \neq v\}$. As we consider only graphs defined over the same set of vertices, in the rest of this paper each graph is simply identified by its set of edges $E \subseteq V^2$. The set of all possible graphs defined over $V$ is then the powerset of  $V^2$, that we denote as $\mathbb{G}(V) = 2^{V^2}$.
Given two groups of such graphs, a condition group (class 1) and a control group (class 0), a graph classification model can be trained to discriminate among the two classes, i.e., to predict the class for a new graph $E \in  \mathbb{G}(V)$, by using any graph classification method.

\spara{Problem statement.}
We are given a \emph{black-box graph classifier}, denoted as a function $f: \mathbb{G}(V) \rightarrow \{0,1\}$. We assume that we do not know anything about $f$ and we can only query it as an oracle, i.e., giving it a graph in input and getting as output the predicted class.
Given a specific graph $E \in \mathbb{G}(V)$, a \emph{counterfactual graph}  is another graph $E' \in \mathbb{G}(V)$, such that $f(E') = 1 - f(E)$.  As discussed in Section \ref{sec:intro}, we are not interested in just any counterfactual instance, but we want to find a $E'$ that is as close as possible to $E$.  As distance measure we adopt a simple \emph{edit distance}: $d(E,E') = | E \Delta E'|$  where $\Delta$ denotes the \emph{symmetric difference} of the two sets, i.e.,  $E \Delta E' = (E \setminus E') \cup (E' \setminus E)$. This is the set of edges that need~to~be removed and the edges that need to be added to go from $E$~to~$E'$.

\smallskip

\begin{mdframed}[innerbottommargin=2pt,innertopmargin=2pt,innerleftmargin=3pt,innerrightmargin=3pt,backgroundcolor=gray!10,roundcorner=10pt]
\begin{problem}[Counterfactual Graph Search]\label{prob:counterfactual_search}
Given a black-box graph classifier $f: \mathbb{G}(V) \rightarrow \{0,1\}$ and a graph $E \in \mathbb{G}(V)$, find a counterfactual graph $E^* \in \mathbb{G}(V)$ such that:
$$
E^* = \argmin_{\substack{E'  \in \mathbb{G}(V):\\ f(E') = 1 - f(E)}} d(E,E').
$$
\end{problem}
\end{mdframed}

As discussed previously, having a counterfactual graph which is as close as possible to the original graph, allows to produce a concise explanation, as \emph{the edit distance corresponds exactly to the dimension of the explanation}.

Given that the search space has size $|\mathbb{G}(V)| = 2^{\frac{n(n-1)}{2}}$ where $n = |V|$,
thus exponential in the number of vertices, and that we do not know anything about the black-box classifier $f$, we cannot hope for a polynomial-time algorithm that solves Problem \ref{prob:counterfactual_search} exactly. What we can do is heuristically explore the search space, querying the black-box classifier $f$ as an oracle over new instances $E' \in \mathbb{G}(V)$.
The number of calls to the oracle $f$ is an important measure when assessing search strategies, as different black-box classifiers might have different time requirements to produce the classification for a new instance $E'$. Therefore, in our framework, we will always generate the closest counterfactual that we can find during a search with a number of calls to the oracle $f$, limited by a user-defined parameter $\eta \in \mathbb{N}^+$.

\section{Counterfactual Graph Search}
\label{sec:algorithms}
As we can not hope to have a polynomial-time exact algorithm for \textsc{Counterfactual Graph Search},
we have tried several simple heuristic-search methods, all having comparable performances. In the rest of this section we present one of these approaches:
 a \emph{bidirectional local-search heuristic} which, while being simple, provides good performance as we will show in Section \ref{sec:experiments}.

In the first phase, we move away from the original graph $E$ looking for a first counterfactual graph $E_c$.
%
%

In the second phase, we iteratively try to produce a better counterfactual, by adding or removing one or more edges from a pool of candidates, i.e., the symmetric difference of the original graph $E$ and the current best counterfactual $E_c$.
We propose two main variants of this general heuristic search schema.
\begin{description}
  \item[Oblivious:] the algorithm is only given the starting graph $E$ and the possibility of querying the black-box.
  \item[Data-driven:] the algorithm has access to a dataset of brain networks $D$, which could be, e.g., the training set used for building the black-box, a subset or a superset of it.
\end{description}
We next describe in more details our heuristic search approaches.

\subsection{Phase 1: Finding the first counterfactual}\label{subsec:alg:phase1}
\spara{\textsf{Oblivious Forward Search}.} As we have access only to the given graph $E$ and the black-box, all we can do is iteratively edit $E$ (adding or removing edges) until we produce a counterfactual graph.  The pseudocode of this procedure, dubbed \textsf{Oblivious Forward Search} (\ofs), is given in Algorithm \ref{alg:OFS}.

More in details, at each step (lines 2 to 8) we select $k$ edges to be changed: either edges in $V^2 \setminus E$ to be added to $E$, or edges in $E$ to be removed. As the former pool of candidate is typically much larger than the latter one, in order to avoid biasing the search in favor of edge additions over removal, before each edit operation we flip an unbiased coin ($rnd()$ line 6) to decide whether to add (lines 6--7) or remove (line 8) an edge. Then one edge $e$ to be changed is picked from the corresponding pool of candidates uniformly at random ($pick$ function). A list $L$ of already used edges is maintained, to avoid adding and removing the same edge multiple times.

\begin{algorithm}[h]
	\caption{\textsf{Oblivious Forward Search} (\ofs)} \label{alg:OFS}
	\begin{algorithmic}[1]
		\REQUIRE $E \in \mathbb{G}(V), \eta \in \mathbb{N}^+, k \in \mathbb{N}^+$, black-box $f: \mathbb{G}(V) \rightarrow \{0,1\}$
		\ENSURE $E_c \in \mathbb{G}(V)$ such that $f(E_c) = 1 - f(E)$ or fail
\STATE $i \leftarrow 0; E' \leftarrow E$; $L \leftarrow \emptyset $
\WHILE{$f(E') = f(E)$ \textbf{and} $i < \eta$}		
\STATE $i$++; $j \leftarrow 0$
\WHILE{$j < k$}		
\STATE $j$++
\STATE \textbf{if} $rnd() < 0.5$ \textbf{then} $e \leftarrow  pick(1, V^2 \setminus (E \cup L));$
\STATE $\;\;\;\;\;\;\;\;\;\;\;\;\;\;\;\;\;\;\;\;\;\; E' \leftarrow E' \cup \{e\}; L \leftarrow L \cup \{e\}$;
\STATE \textbf{else} $e \leftarrow pick(1,E \setminus L); E' \leftarrow E' \setminus \{e\}; L \leftarrow L \cup \{e\}$;
\ENDWHILE
\ENDWHILE
\STATE \textbf{if} $f(E') = 1 - f(E)$ \textbf{then return} $E'$ \textbf{else fail}
\end{algorithmic}
\end{algorithm}

The search is called ``forward'' because it starts from the original graph $E$ and it moves away from it. In particular, at each step, the edit distance $d(E',E)$ between the current graph $E'$ and the original graph $E$ increases of $k$ edges.

\spara{\textsf{Data-driven Forward Search}.} We next consider the case in which we have access to a database of brain networks  $D = \{(E_i,y_i)\}_{i=1}^n$ s.t. $E_i \in \mathbb{G}(V), y_i \in \{0,1\}$, which could be, e.g., the publicly available training set used for building the black-box, a subset or a superset of it, or even a different dataset but having the same structure of the training set, with a condition and a control group for the same mental disease. In this setting, we can use the distribution of edges among the two classes to bias the random selection (i.e., the $pick$ function at lines 6 and 8 of Algorithm \ref{alg:OFS}), \emph{to favor edges which are better at discriminating between the two classes}. Specifically, we define a weighting function $w: V^2\rightarrow \mathbb{Z}$ that assigns a weight to each edge based on its occurrences in the two classes as follows.
For each $e \in V^2$ we define $D^+(e) = \{ (E_i,y_i) \in D | e \in E_i \wedge y_i = f(E)\}$ the set of graphs in the database that contain $e$ and have label concordant with the classification of the original graph $f(E)$, and similarly, we define $D^-(e) = \{ (E_i,y_i) \in D | e \in E_i \wedge y_i = 1 - f(E)\}$ the set of graphs in the database that contain $e$ and have label discordant with the classification of the original graph.

Then we define:
$$
w(e) = \begin{cases}
|D^+(e)| - |D^-(e)| & \text{if} \; e \in E  \\
|D^-(e)| - |D^+(e)| & \text{if}  \; e \in V^2 \setminus E  \\
\end{cases}
$$

The idea is that we want to remove edges from $E$ that are strongly characterizing of the class $f(E)$, so to more easily produce a counterfactual with small edit distance, or alternatively, add to $E$ edges that are strongly characterizing of the opposite class $1-f(E)$.

Therefore, in the setting in which we have access to $D$,
we substitute  $pick()$ at lines 6-8 of Algorithm \ref{alg:OFS} with a function that picks edges with probability proportional to $max(\epsilon,w(.))$, where $\epsilon$ is a very small positive constant, used to avoid null probabilities while making the edges with negative weights unlikely to be picked. The resulting method is called \textsf{Data-driven Forward Search} (\dfs).

\subsection{Phase 2: Finding a better counterfactual}
In the second phase, we start from the first counterfactual graph produced by phase 1, let us denote it $E_c^1$, and we go back towards the original graph $E$. We do so by randomly picking edges in the symmetric difference $E \Delta E_c^1$ and modifying $E_c^1$ consequently. The approach is justified by the key observation that modifying $E_c^1$ by adding or removing edges from the symmetric difference with $E$,  we obtain a graph $E'$ which is guaranteed to have a smaller edit distance from $E$, i.e., $d(E',E) < d(E_c^1,E)$.

\spara{\textsf{Oblivious Backward Search}.} Our heuristic search method reported in Algorithm \ref{alg:OBS}, adapts the number of edges $k$ that are changed in each iteration, so to try to reach the classification border faster. If in one iteration $i$, the candidate
graph $E^i_c$ turns out to be a counterfactual (line 7), we increase $k$ (line 8) as we are not yet at the classification border, if instead, it is not a counterfactual graph (meaning that the backward search has crossed the classification border), we reduce the value of $k$ for the next iteration (line 9). At every iteration, $E_c$ keeps the current best counterfactual graph, while $E_d$ keeps the symmetric difference, i.e., the pool from which to select the edges to add or remove.

\begin{algorithm}[h]
	\caption{\textsf{Oblivious Backward Search} (\obs)} \label{alg:OBS}
	\begin{algorithmic}[1]
		\REQUIRE $E,E^{1}_c \in \mathbb{G}(V), \eta, k \in \mathbb{N}^+$, black-box $f: \mathbb{G}(V) \rightarrow \{0,1\}$
	\ENSURE $E_c \in \mathbb{G}(V)$ such that $f(E_c) = 1 - f(E)$
    \STATE $E_c \leftarrow E^{1}_c; i \leftarrow 0$;
     \STATE $E_d \leftarrow E \Delta E_c$
    \WHILE{$i < \eta$ \textbf{and} $|E_d| > 0$}
    \STATE $i$++; $k \leftarrow \min(k, |E_d|)$;	
	\STATE $E^{k}_d \leftarrow pick(k,E_d)$ 
	\STATE $E^{i}_c \leftarrow (E_c \cup (E^{k}_d \setminus E_c)) \setminus (E^{k}_d \cap E_c)$
    \IF{$f(E^{i}_c) = 1 - f(E)$}	
    \STATE $k$++; $E_c \leftarrow E^{i}_c; E_d \leftarrow E \Delta E_c$
    \STATE \textbf{else if} $k > 1$ \textbf{then} $k$- -		
    \STATE $\quad\quad\quad\quad\quad\;$ \textbf{else} $E_d \leftarrow E_d \setminus E^{k}_d$;
\ENDIF
\ENDWHILE
\STATE \textbf{return} $E_c$
\end{algorithmic}
\end{algorithm}

In line 5, the function $pick(k,E_d)$ picks uniformly at random $k$ edges from the pool of candidate and then add or remove them from $E_c$ accordingly (line 6) to produce the candidate graph $E_c^i$.
When $k$ reaches the value of 1, we stop decreasing $k$ and start removing the edges that have been already tested from the pool of candidate edges $E_d$: as we are trying edges one by one, it does not make sense to try them multiple times. Eventually, when $E_d$ is emptied, the algorithm terminates (line 3). The other termination condition is when the algorithm has reached the maximum allowed number $\eta$ of calls to the black-box.

\spara{\textsf{Data-driven Backward Search}.} A key part of the second phase of counterfactual graph search is the selection of the edges to change at each iteration, which, in the oblivious case, is implemented by a uniform random selection (function $pick(k,E_d)$, in line 6 of Algorithm \ref{alg:OBS}).
In the case in which we have access to a database of brain networks  $D = \{(E_i,y_i)\}_{i=1}^n$, similarly to what done for \dfs, we can use the distribution of edges among the two classes to bias the random selection, favoring edges which are better at discriminating between the two classes.
In this case we define the weight as:
$$
w(e) = \begin{cases}
|D^+(e)| - |D^-(e)| & \text{if} \; e \in E \setminus E_c    \\
|D^-(e)| - |D^+(e)| & \text{if}  \; e \in E_c \setminus E  \\
\end{cases}
$$
where $D^+(e)$ and $D^-(e)$ are defined as in \S\ref{subsec:alg:phase1}.
By substituting the function $pick(k,E_d)$, in line 5 of Algorithm \ref{alg:OBS} with a function that picks edges from $E_d$ with probability proportional to $max(\epsilon,w(.))$, where $\epsilon$ is a very small positive constant, we obtain the method called \textsf{Data-driven Backward Search} (\dbs).

\begin{algorithm}[h]
\caption{Baseline: \textsf{Dataset Search} (\ds)}\label{alg:DS}
\begin{algorithmic}[1]
\REQUIRE $E \in \mathbb{G}(V)$, black-box $f: \mathbb{G}(V) \rightarrow \{0,1\}$,\\ dataset $D = \{(E_i,y_i)\}_{i=1}^n$ s.t. $E_i \in \mathbb{G}(V), y_i \in \{0,1\}$
\ENSURE $E_c \in \mathbb{G}(V)$ such that $f(E_c) = 1 - f(E)$ or fail
\STATE $E_c \leftarrow \emptyset; \delta \leftarrow |V^2|$
\FORALL{$(E_i, y_i) \in D$ s.t. $y_i = 1 - f(E)$}
\IF{$d(E,E_i) < \delta$}
\IF{$f(E_i) = 1 - f(E)$}
\STATE $E_c \leftarrow E_i; \delta \leftarrow d(E,E_i)$
\ENDIF
\ENDIF
\ENDFOR
\STATE \textbf{if} $E_c \neq \emptyset$ \textbf{then return} $E_c$ \textbf{else fail}
\end{algorithmic}
\end{algorithm}
\subsection{A simple baseline}
When we have access to the dataset $D$ we can also consider a baseline that simply searches in $D$ for the counterfactual graph with the minimum edit distance from the original graph $E$.
This baseline, called \textsf{Dataset Search} (\ds) and described in Algorithm \ref{alg:DS}, is interesting because it answers the question of finding the \emph{best counterfactual graph among the real networks} available.

As we have access to the real labels of the graphs in $D$, we avoid checking those graphs which have the same class label as $E$ (line 2). For the other ones, we first check the edit distance from $E$, and only if it is smaller than the current best one $\delta$ (line 3), we call the black-box (line 4). If the graph is in the counterfactual class and it is correctly classified so by the black-box, then it becomes the current best counterfactual graph (line 5).

\medskip
\section{Assessment on Brain Networks}
\label{sec:experiments}
In this section we assess the performance of the counterfactual graph search methods introduced in Section \ref{sec:algorithms}.

\subsection{Experiment setting}
\spara{Human-brain datasets.}
Our experiments are performed on two publicly-available datasets.
The first dataset, about \emph{Autism Spectrum Disorder} (\textsf{ASD}), is taken from the Autism Brain Imagine Data Exchange (ABIDE)\footnote{\url{http://fcon_1000.projects.nitrc.org/indi/abide/}} \cite{craddock2013abide}.
In particular, we focus on the portion of dataset containing children below 9 years of age \cite{lanciano2020}. These are 49 individuals in the condition group, labeled as \emph{Autism Spectrum Disorder} (\textsf{ASD}) and 52 individuals in the control group, labeled as \emph{Typically Developed} (\textsf{TD}).
The second dataset, about  \emph{Attention Deficit Hyperactivity Disorder} (\textsf{ADHD}), is taken
from the USC Multimodal Connectivity Database (USCD)\footnote{\url{http://umcd.humanconnectomeproject.org/umcd/default/browse_studies}} \cite{adhd}, in particular from the ADHD200\_CC200 study. This contains 190 individuals in the condition group, labeled as \textsf{ADHD} and 330  individuals in the control group, labeled as \textsf{TD}.
Both datasets study brain functional connectivity by means of functional magnetic resonance imaging (fMRI), that allows to measure the statistical association between ROIs, as reviewed in \S\ref{sec:preliminaries}. For the parcellation of the brain we use the AAL \cite{aal} atlas for the \textsf{ASD} dataset  ($|V| = 116$) and the Craddock 200 (CC200) \cite{craddock2012threshold} for the \textsf{ADHD} dataset ($|V| = 190$).
The pre-processing needed to go from the time series to the correlation matrices is performed in accordance with the literature.\footnote{See \url{http://preprocessed-connectomes-project.org/abide/dparsf.html} for \textsf{ASD} dataset and \url{https://ccraddock.github.io/cluster_roi/atlases.html} for \textsf{ADHD} dataset.}

The final data-preparation step is to go from correlation matrices to networks.
In accordance with large portion of the literature, e.g., \cite{p1,p2,lanciano2020}, we transform
 a correlation matrix $C=\{c_{ij}\}$  into an adjacency matrix $A=\{a_{i,j}\}$ by setting
  $a_{ij}=1$ when $c_{ij}$ is larger than a given threshold, and setting  $a_{ij}=0$ otherwise.
The threshold is selected for both datasets as the $90th$ percentile of the distribution of the correlation matrix values.


\spara{Black-box classifiers.} Our method is totally agnostic of the black-box classifier under analysis:
it just sees the black-box as a function $f: \mathbb{G}(V) \rightarrow \{0,1\}$, i.e., an oracle that can be queried with a graph and which gives back a label in $\{0,1\}$. As such the specific black-box we use is not so important in our empirical assessment. Nevertheless, for variety sake, we use 4 different black-box classifiers for our brain network context: \textsf{Contrast Subgraphs}~\cite{lanciano2020},
\textsf{Graph2Vec} \cite{annamalai2017graph2vec},
\textsf{Sub2Vec} \cite{adhikari2018sub2vec}, and
\textsf{Autoencoders} \cite{gutierrez2019embedding}.
These methods are very diverse and cover a wide spectrum of complexity for the explanation task.
Each of these methods (the details about their parameters settings are provided in Appendix) produces a vector for each graph in the training set. Then a $k$-nearest neighbors (KNN) or a Support Vector Machine (SVM) is adopted to produce the final classification.

\subsection{Methods assessment} \label{subsec:exp:comparison}
We first assess the counterfactual graph search algorithms w.r.t. (1) the quality of the counterfactual produced, measured in terms of edit distance from the original graph, and (2) the number of calls to the black-box. We run an experiment for each graph in the training set. Each experiment is run 5 times and we report average results over the 5 runs. For each phase of the counterfactual graph search, we allow a maximum number of calls to the black-box $\eta = 2000$. We set $k = 5$ in all experiments.

\begin{figure}[t!]
   \centering
\begin{tabular}{cc}
\textsf{\small Phase 1} & \textsf{\small Phase 1 and Phase 2}\\
	\includegraphics[width=0.215\textwidth]{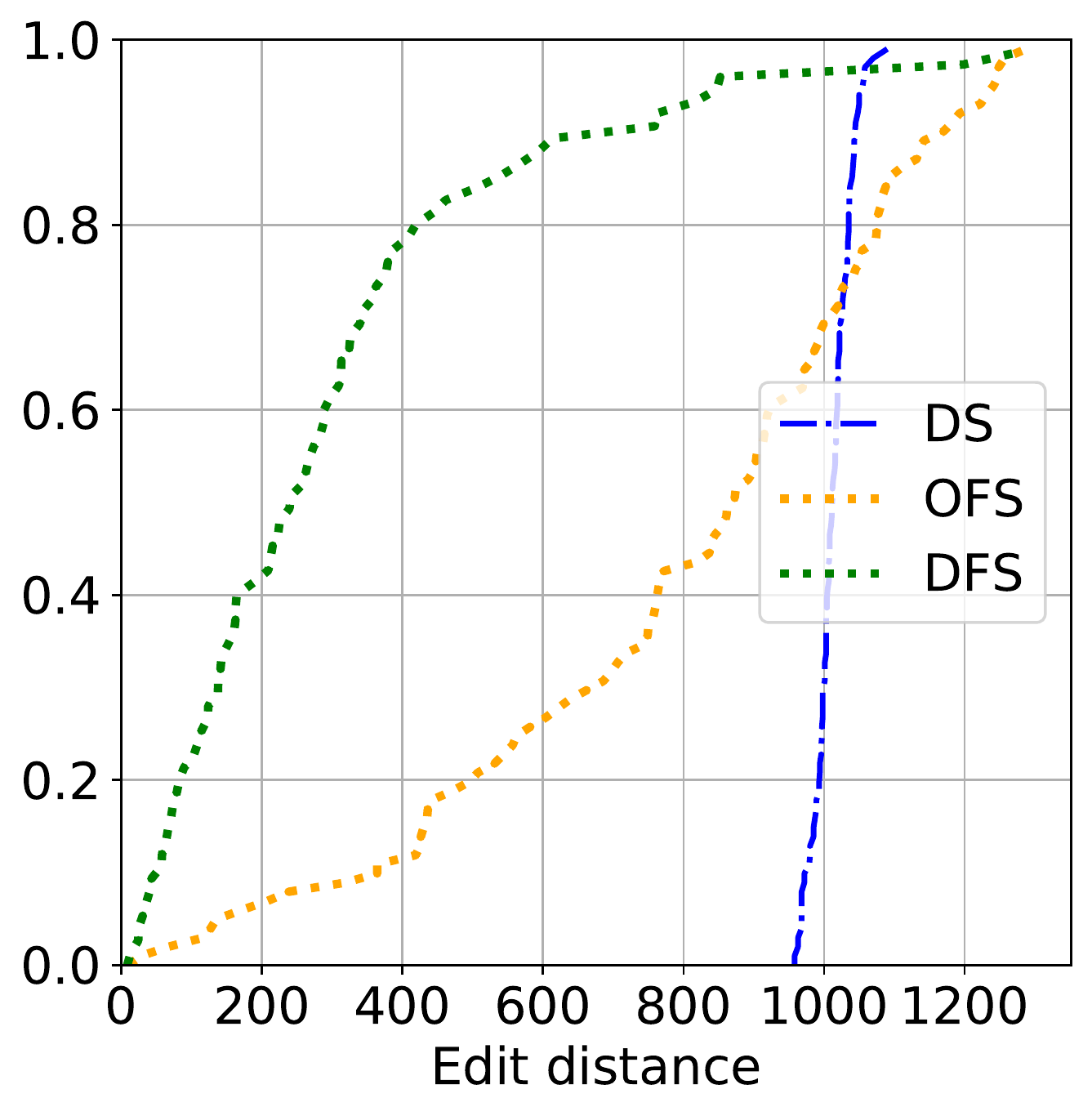}&
	\includegraphics[width=0.22\textwidth]{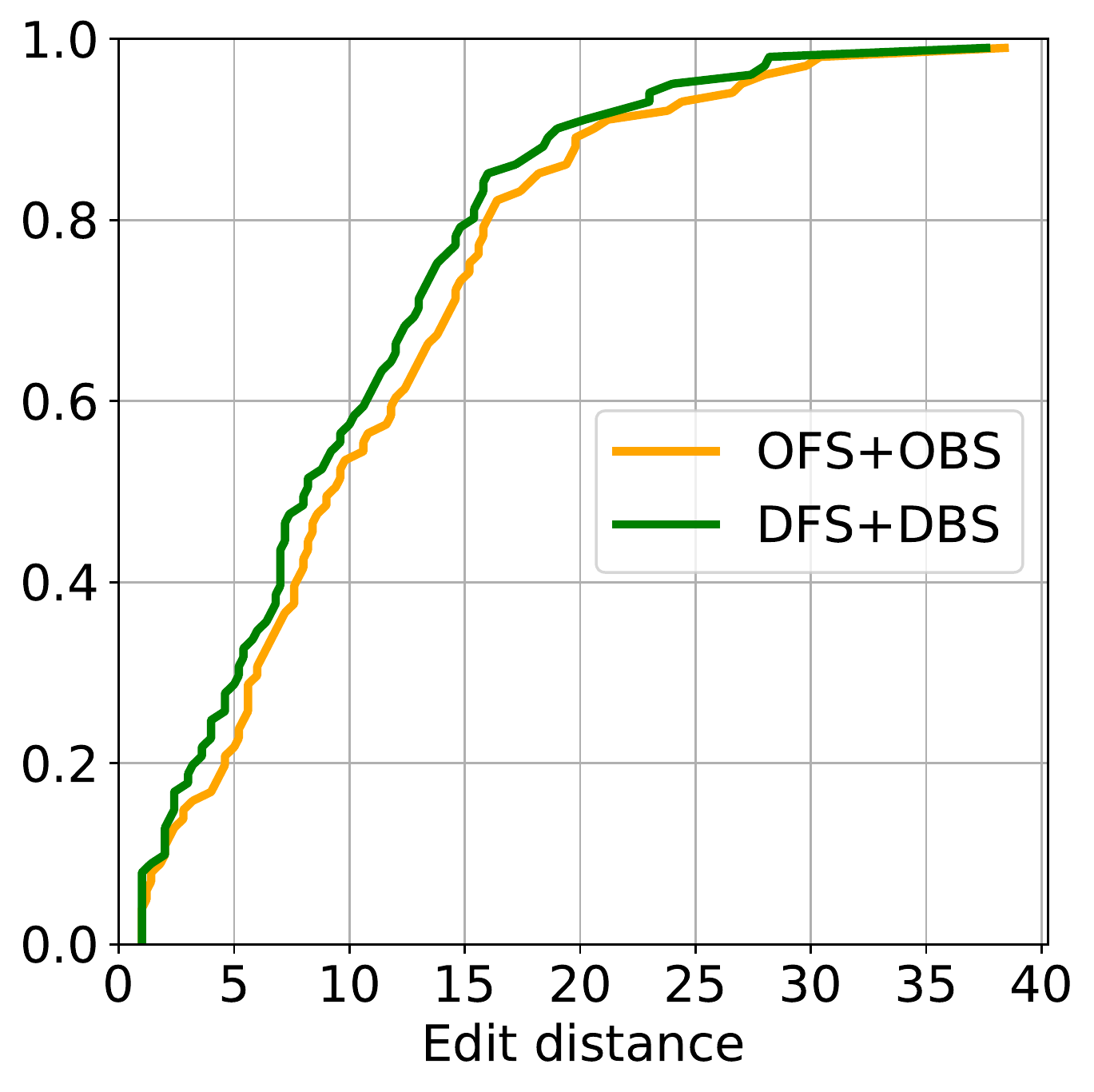}\\
\end{tabular}
	\vspace{-4mm}
\caption{Cumulative distribution of $d(E,E_c)$ between the input graph $E$ and its counterfactual $E_c$, on the \textsf{ASD} dataset using \textsf{Contrast Subgraphs + SVM} as black-box. One experiment  (average over 5 runs) for each graph in the dataset.}
	\label{fig:methods1}
\end{figure}

Figure \ref{fig:methods1} reports the cumulative distribution of the edit distance between the input graph and its counterfactual, on the \textsf{ASD} dataset using as black-box \textsf{Contrast Subgraphs + SVM}.
The left plot reports the distribution after the first phase and includes the baseline \textsf{Dataset Search} (\ds). We can observe that, as expected, the data-driven approach \dfs\ is much more effective than the oblivious \ofs\ in finding the first counterfactual. In 80\% of the cases, the edit distance of the first counterfactual is below 400, while the average counterfactual distance in the dataset is approximately 1000.

\begin{figure}[t!]
   \centering
\includegraphics[width=0.3\textwidth]{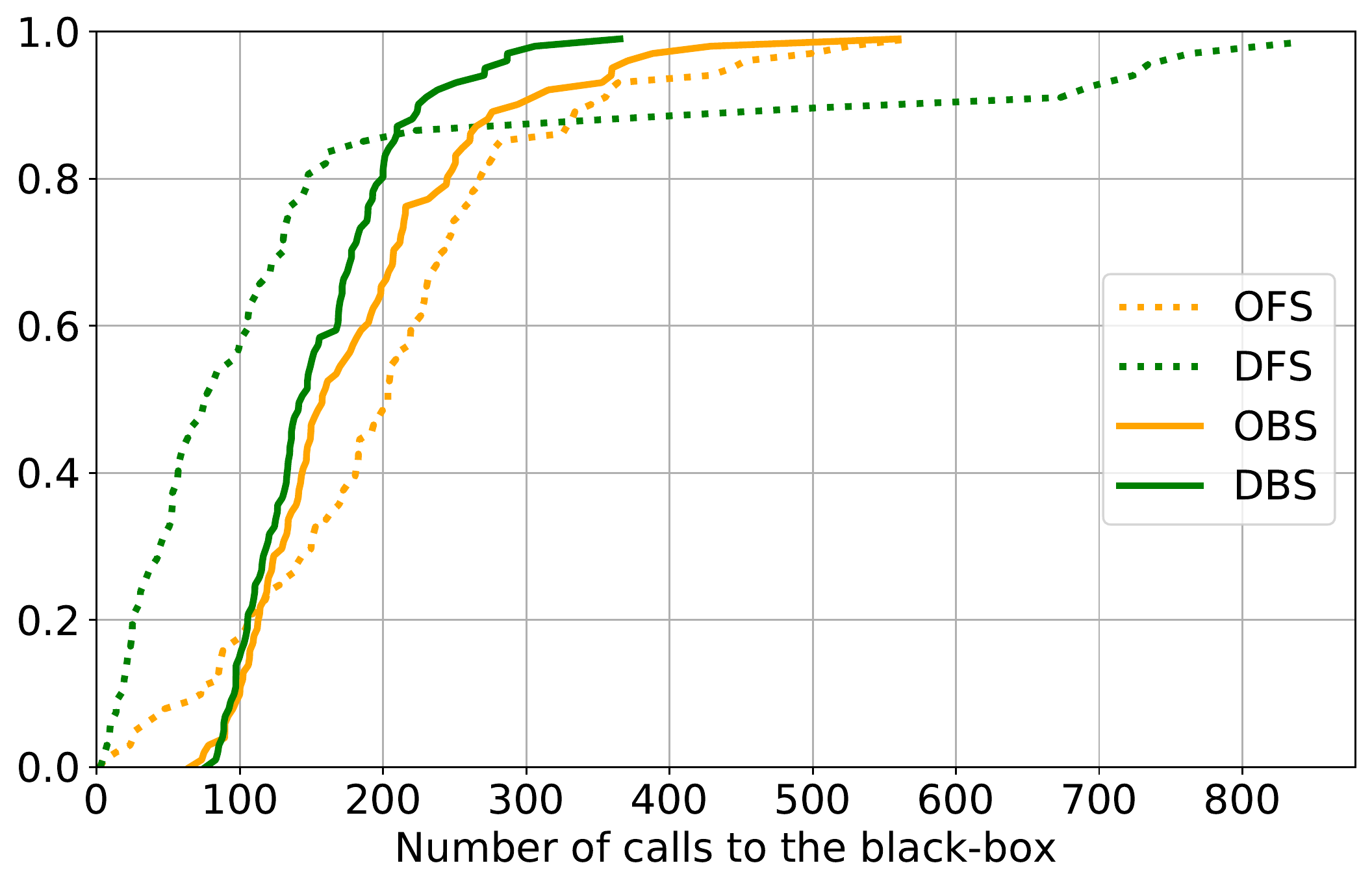}
\vspace{-2mm}
\caption{Cumulative distribution of the number of calls to the black-box (same setting as Figure \ref{fig:methods1}).}
	\label{fig:methods2}
\end{figure}

However, after phase 2, the difference in quality between the data-driven approach and the oblivious one get much smaller, indicating that \emph{both approaches converge very close to the best possible counterfactual graph}. A large difference between the oblivious and the data-driven case is instead observable in the number of calls to the black-box, which is reported in Figure \ref{fig:methods2}.

From this preliminary assessment, we conclude that: (1) both oblivious and data-driven approaches produce similar results in term of quality, but the data-driven case converges to a good counterfactual graph faster (with less calls to the black-box); (2) the counterfactual graph produced by the proposed methods is much better in quality than what we could find by searching among the real graphs in the dataset.

In Table \ref{tab:agnosticmethod} we report the same statistics for all the black-boxes and on both datasets, using only the oblivious counterfactual search approach (\ofs+\obs).
We can observe that, when dealing with black-boxes that explicitly keep in consideration the node identity awareness property of brain networks (i.e., \textsf{Contrast Subgraphs}~\cite{lanciano2020} and \textsf{Autoencoders} \cite{gutierrez2019embedding}), we typically get counterfactual graphs with smaller edit distance than the other two models. It is worth mentioning that in average, the best counterfactuals
that one can obtain by using real examples in the dataset (i.e., by means of \ds\ baseline), have edit distance from their original graphs of approximately 1000 in the \textsf{ASD} dataset and 2250 in the  \textsf{ADHD} dataset.


While there is no shared consensus on how to evaluate post-hoc explanations, when counterfactuals come into play we argue that the smaller the edit distance the better the explanation. When counterfactuals are systematically closer to the original data points, the black-boxes are easier to interpret. We can use the minimum edit distance as a measure to select among different classification methods, in domain such as brain networks in which post-hoc explainability is a selection criterion as important as the accuracy.

\subsection{Assessment against a white-box} \label{subsec:exp:whitebox}
Since, in general, it is not possible to compute the optimal counterfactual exactly, our methods are heuristic in nature. It is thus interesting to compare our counterfactual solution with the optimal counterfactual, in a setting in which an optimal counterfactual can be identified.

\begin{figure}[t!]
    \centering
    \includegraphics[width=\linewidth]{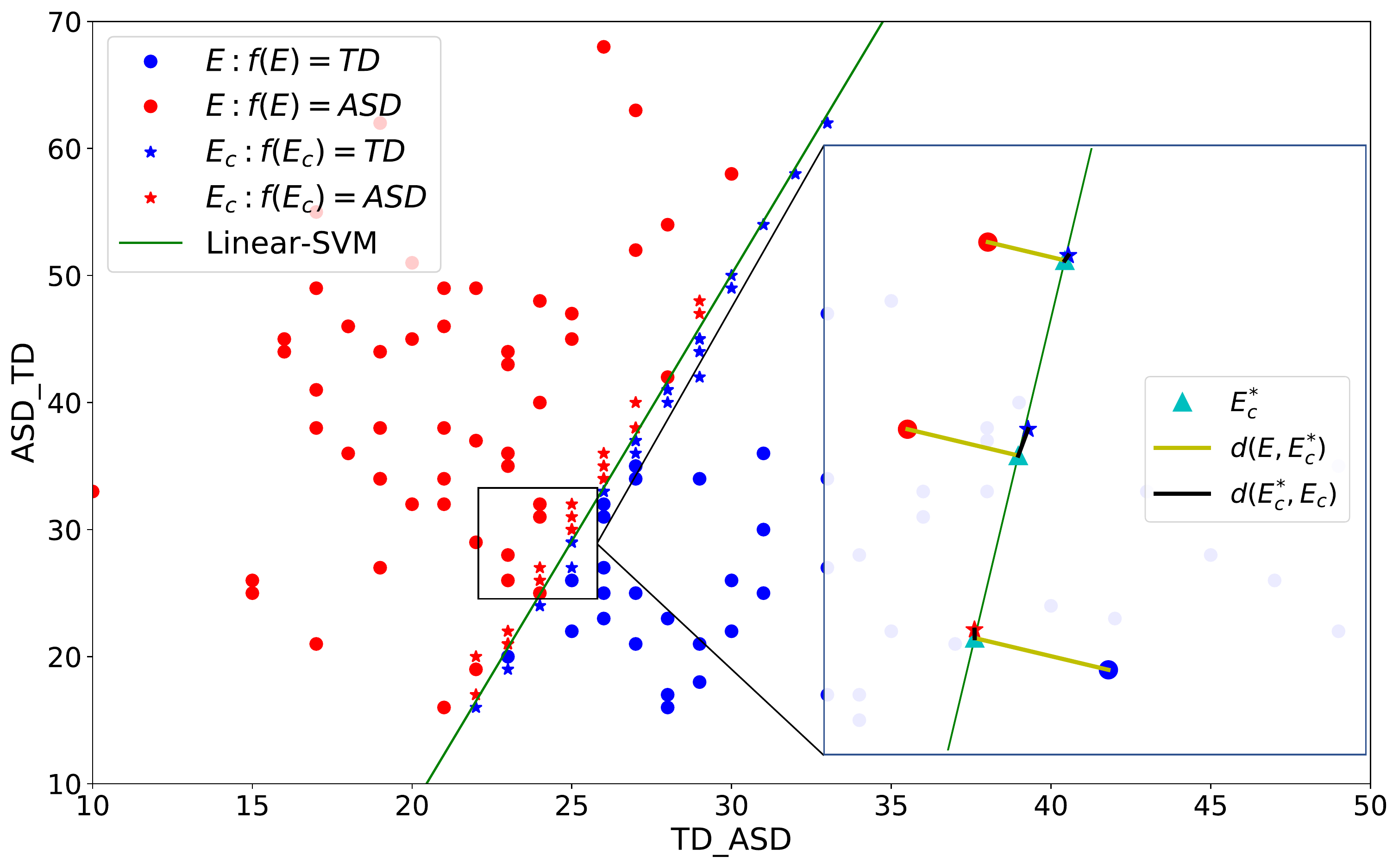}
    \vspace{-6mm}
    \caption{Real setting (\textsf{ASD} dataset): using \textsf{Contrast Subgraphs} to produce a 2-dimensional embedding on which we learn a linear classifier (green line) by means of SVM. For each individual $E$ (dot) we produce one counterfactual $E_{c}$ (star) using \ofs+\obs\ with $\eta = 2000$ and $k = 5$. On the right-hand side we show a zooming on the classification border where we can appreciate how close the produced counterfactuals are to the optimal ones $E_c^*$ (cyan triangles).}
    \label{fig:optimumcount}
\end{figure}

We achieve this by means of a very simple white-box, i.e., a classifier for which we can see the inner logic. In particular, we train a linear classifier built on a 2-dimensional embedding. This gives us a cartesian $(x,y)$-plane as the one in Figure~\ref{fig:optimumcount}, in which it is possible to identify the optimal counterfactual \emph{geometrically}.
In fact, if we consider the original graph $E$ as a point $(x_o,y_o)$ on the $(x,y)$-plane and the classifier $f$ as a line $r_f: y =mx+c$, we can obtain the optimal counterfactual graph $E^*_{c}$ as the point $(x_c,y_c)$ as follows:
\begin{enumerate}
	\item Generate the line $r_p$ perpendicular to the classifier line $r_f$ and passing through $(x_o,y_o)$, with equation $y=-\frac{1}{m}(x-x_o)+y_o$;
	\item get the point $(x_c,y_c)$ obtained by the intersection of $r_p$ with $r_f$, with the following coordinates $x_c=\frac{(x_o+m y_o-m c)}{m^{2} +1}$ and $y_c=\frac{m(x_o+m y_o-m c)}{m^{2} +1}+c$.
\end{enumerate}

\begin{table}[t!]
\centering
\caption{Counterfactual graph search statistics on for all the black-boxes and on both datasets, using only the oblivious counter-factual search approach (\ofs+\obs). Same experimental settings as before: One experiment  (average over 5 runs) for each graph in the dataset, $\eta = 2000$  maximum calls to the black-box for each phase of the counterfactual search. We report the 10th, 25th, 50th, 75th, and 90th percentiles.}
\vspace{-2mm}
\begin{adjustbox}{width=1\linewidth}
\begin{tabular}{r|c|c|c|c|c|c|c|c|c|c|c|}
  \multicolumn{1}{c}{}& \multicolumn{5}{c}{\textbf{\textsf{ASD}:} Edit Distance} & \multicolumn{1}{c}{$\;$}& \multicolumn{5}{c}{\textbf{\textsf{ASD}:} Calls to Oracle} \\
 \cline{2-6}  \cline{8-12}
 & 10-p & 25-p & 50-p & 75-p & 90-p & & 10-p & 25-p & 50-p & 75-p & 90-p\\
 \cline{2-6}  \cline{8-12}
 \textsf{Cont. Sub.} & 2 & 5.4 & 9 & 15.2 & 19.8 & & 100 & 119.2 & 157.4 & 214.6 & 276.2 \\
 \textsf{Sub2Vec } & 4 & 47.5 & 134 & 285 & 476 & & 110.5 & 175.5 & 299.5 & 492 & 617.5 \\
 \textsf{Graph2Vec} & 1.5 & 6.8 & 33.5 & 119 & 207.4 & & 112.2 & 131 & 181.6 & 346.3 & 445\\
 \textsf{Autoenc.} & 1.6 & 2.4 & 3.8 & 6 & 9.2 & & 82.2 & 101 & 141 & 209.4 & 275.4 \\
 \cline{2-6}  \cline{8-12}
 \multicolumn{12}{c}{$\;$}\\
  \multicolumn{1}{c}{}& \multicolumn{5}{c}{\textbf{\textsf{ADHD}:}  Edit Distance} & \multicolumn{1}{c}{$\;$}& \multicolumn{5}{c}{\textbf{\textsf{ADHD}:} Calls to Oracle} \\
 \cline{2-6}  \cline{8-12}
 & 10-p & 25-p & 50-p & 75-p & 90-p & & 10-p & 25-p & 50-p & 75-p & 90-p\\
 \cline{2-6}  \cline{8-12}
  \textsf{Cont. Sub.} & 9 & 20.6 & 43.8 & 70 & 100.2 & & 173.2 & 249.2 & 406 & 578.2 & 764.7 \\ %
 \textsf{Sub2Vec} & 8.8 & 56.2 & 185.3 & 367.2 & 742.8 & & 283.88 & 474.5 & 780.56 & 1807.1 & 2000 \\%
 \textsf{Graph2Vec} & 1 & 2.2 & 49.3 & 166 & 411.6 & & 135 & 156.6 & 324.4 & 547 & 1293.2 \\%
 \textsf{Autoenc.} & 1 & 1.8 & 5 & 15.4 & 25.1 & & 129 & 137.8 & 172.5 & 273.8 & 367.5 \\
\cline{2-6}  \cline{8-12}
\end{tabular}
\end{adjustbox}
\label{tab:agnosticmethod}
\end{table}

The  2-dimensional embedding we use in Figure \ref{fig:optimumcount} is obtained by means of the \textsf{Contrast Subgraphs} method~\cite{lanciano2020}, on the \textsf{ASD} dataset. By applying this method we extract a contrast subgraph \textsf{ASD\_TD}, i.e., a set of vertices whose induced subgraph is very dense among \textsf{ASD} individuals and very sparse among \textsf{TD} individuals, and similarly, another
contrast subgraph \textsf{TD\_ASD}, i.e., a set of vertices whose induced subgraph is very dense among \textsf{TD} individuals and very sparse among \textsf{ASD} individuals. Then for each graph in the dataset,
we compute the two dimensions: these are the number of edges in the subgraph induced by the contrast subgraph \textsf{ASD\_TD} ($y$-axis), and the number of edges in the subgraph induced by the contrast subgraph \textsf{TD\_ASD} ($x$-axis). Over this embedding, we use a linear Support Vector Machine (SVM) to build the white-box classifier, represented by the green line in Figure \ref{fig:optimumcount} (which has an accuracy of 0.78 on the training set). Hence we can reconstruct the optimal solution geometrically and measure the error of the counterfactual graph produced (i.e, its distance from the optimal).

As measure of error of the counterfactual search method, we use the average edit distance between the optimal counterfactual graphs $E^*_c$ and the estimated counterfactual graph $E_c$ for all the graphs in the dataset $D$. In the setting reported in Figure \ref{fig:optimumcount} (i.e., \textsf{ASD} dataset,
\textsf{Contrast Subgraphs} + SVM classifier), the average edit distance $d(E^*_c,E_c)$ between the optimal counterfactual graphs $E^*_c$ and the estimated counterfactual graph $E_c$ is 1.83.

\section{Local and Global Explainability}
\label{sec:applications}
\begin{figure}[t!]
    \centering
    \includegraphics[width=\linewidth]{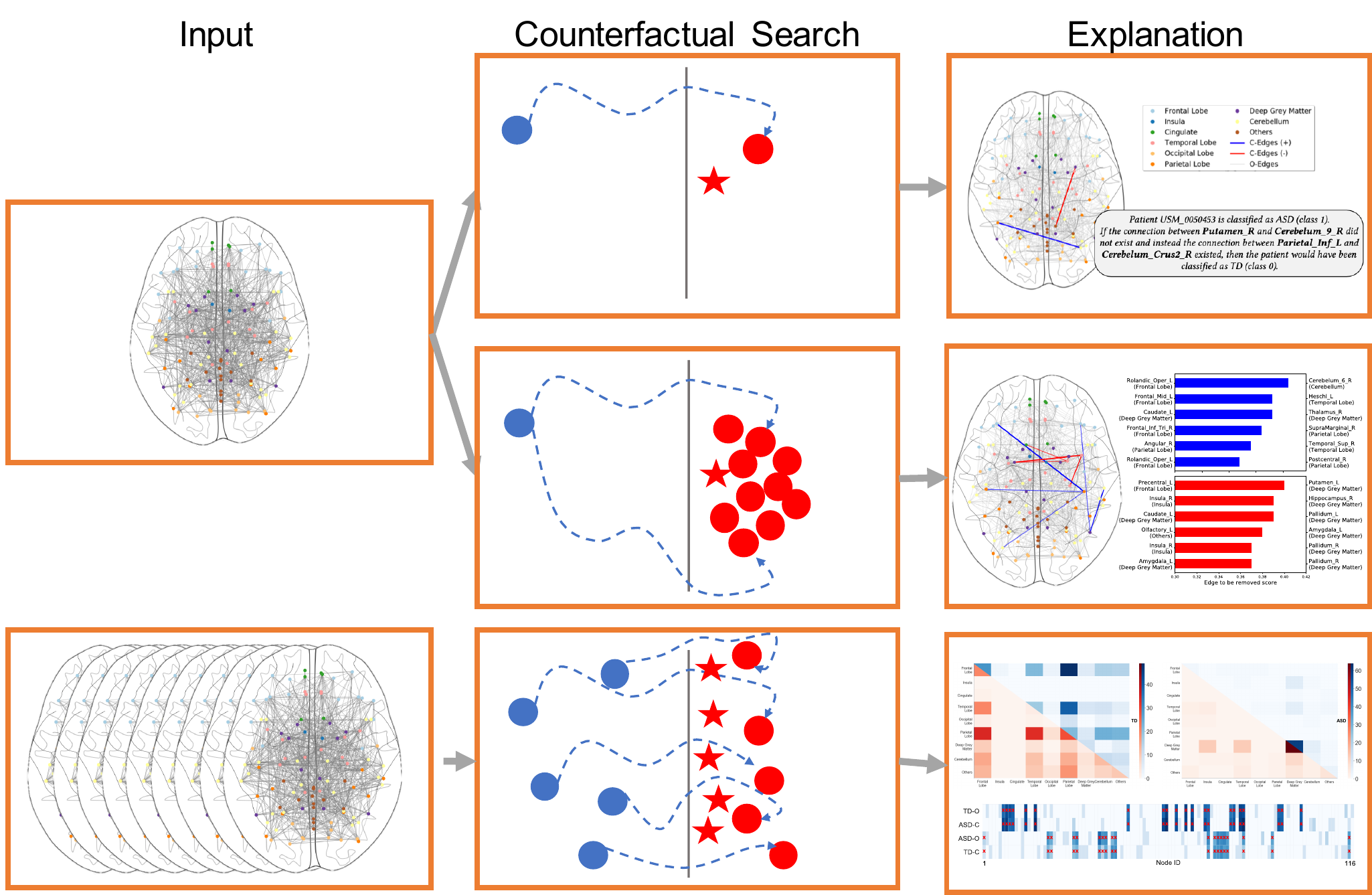}
    \vspace{-4mm}
    \caption{Schematization of the three different types of explanations we can produce by means of graph counterfactuals: Contrastive case-based explanations against counterfactuals (first row), LIME-like (SHAP-like)  local explanations (second row), and global explanations (third row).}
    \label{fig:schema}
\end{figure}

In this section, to showcase the versatility and usefulness of our proposal,  we show how counterfactual graphs can be used as basic building blocks to produce various type of explanations. In particular, we consider three different types of explanations, as summarized in Figure \ref{fig:schema}:

\squishdesc
  \item[\textbf{Contrastive case-based explanations against counterfactuals:}]
  this is the direct presentation of a counterfactual graph as a local post-hoc explanation, as done in the example in Figure \ref{fig:intro} in \S\ref{sec:intro}.

  \item[\textbf{LIME-like (SHAP-like) explanations:}] LIME~\cite{ribeiro2016should}~and~SHAP~\cite{lundberg2017unified} are prominent examples of general explanation methods for classifiers (on tabular data), that explain the prediction for a given example by ranking the input features by their importance in the prediction. We can obtain a similar type of explanation by producing, for a given graph, many counterfactual graphs and collecting statistics about the edges that appear more frequently.

  \item[\textbf{Global explanations:}] finally, by producing counterfactual graphs for many input examples and collecting statistics we can produce meaningful global explanations of any black-box graph classifier.
\squishend

\smallskip

\noindent In the rest of this section we will elaborate further on the latter two points providing concrete examples on our brain network datasets.

\subsection{LIME-like (SHAP-like)  local explanations} \label{subsec:appl:local}
Given a counterfactual graph $E_c$ of a graph $E$, the explanation provided by the counterfactual graph is the symmetric difference $E \Delta E_c = (E \setminus E_c) \cup (E_c \setminus E)$, i.e., the set of edges $E^{c}_{-}= E \setminus E_{c}$ to be removed from $E$, and the set of edges $E^{c}_{+}= E_{c} \setminus E$ to be added to $E$, so to transform the original graph $E$ in its counterfactual graph $E_c$. From this perspective, edges in $E^{c}_{-}$ are edges whose \emph{presence} in $E$ is important to explain the classification $f(E)$.
Similarly, edges in  $E^{c}_{+}$ are edges whose \emph{absence} from $E$ is important to explain $f(E)$.

As for a graph $E$ its counterfactual is not unique, we can produce many counterfactual graphs (say 1000) and collect statistics about the number of times edges appear in $E^{c}_{-}$ or in $E^{c}_{+}$. This statistic has a direct interpretation as importance of the edges in explaining $f(E)$.

\begin{figure}[t!]
    \centering
    \includegraphics[width=1.0\linewidth]{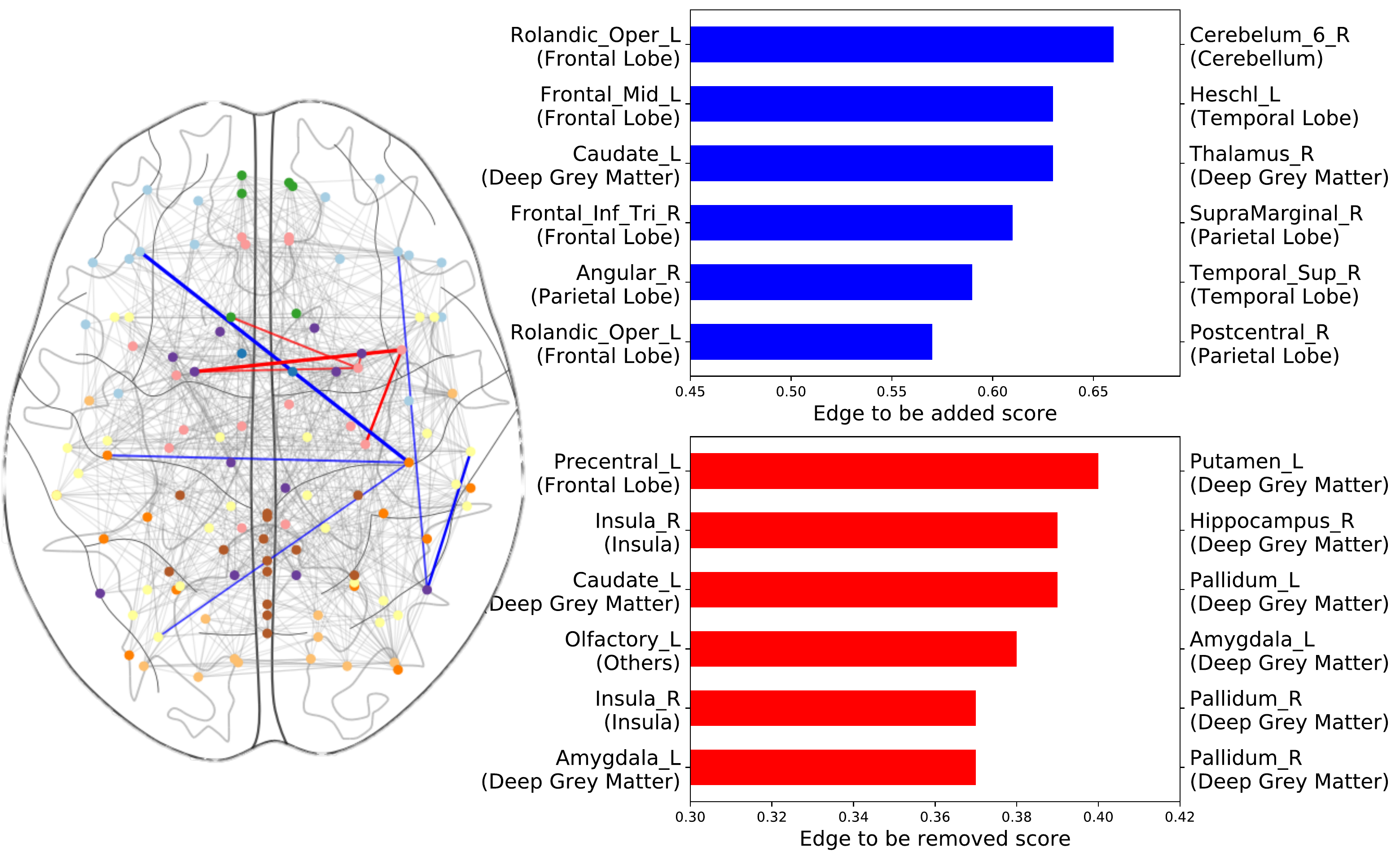}
    \vspace{-4mm}
    \caption{Visualization of LIME-like (SHAP-like) local explanation for the individual \texttt{MaxMun\_d\_0051353} from the \textsf{ASD} dataset, which is classified in the autism class by the classifier of Figure \ref{fig:optimumcount}. In the bar-plots on the right-hand side, we represents the top-6 edges to be added (blue bar) and the top-6 edges to be removed (red bard). These edges are also represented on the brain network (on the left), using the same colors, and with a thickness proportional to their importance.}
    \label{fig:count_edges_local}
\end{figure}

In Figure \ref{fig:count_edges_local} we provide an example of visualization of this local LIME-like (SHAP-like) explanation for a given individual which is classified in the \emph{ASD} class. We can see that, among the edges whose absence is important in explaining the classification of this specific patient, there are many edges in the \emph{frontal lobe} region, while among the edges whose presence is important for this classification, there are many in the \emph{deep grey matter}. These observation are consistent with some neuroscience literature, e.g., \cite{Richards_deep}.


\begin{figure*}[t!]
    \centering
    \begin{tabular}{ccc}
  \textsf{Contrast Subgraphs} &   \textsf{Autoencoders} & \textsf{Graph2vec}\\
    \includegraphics[width=0.32\textwidth]{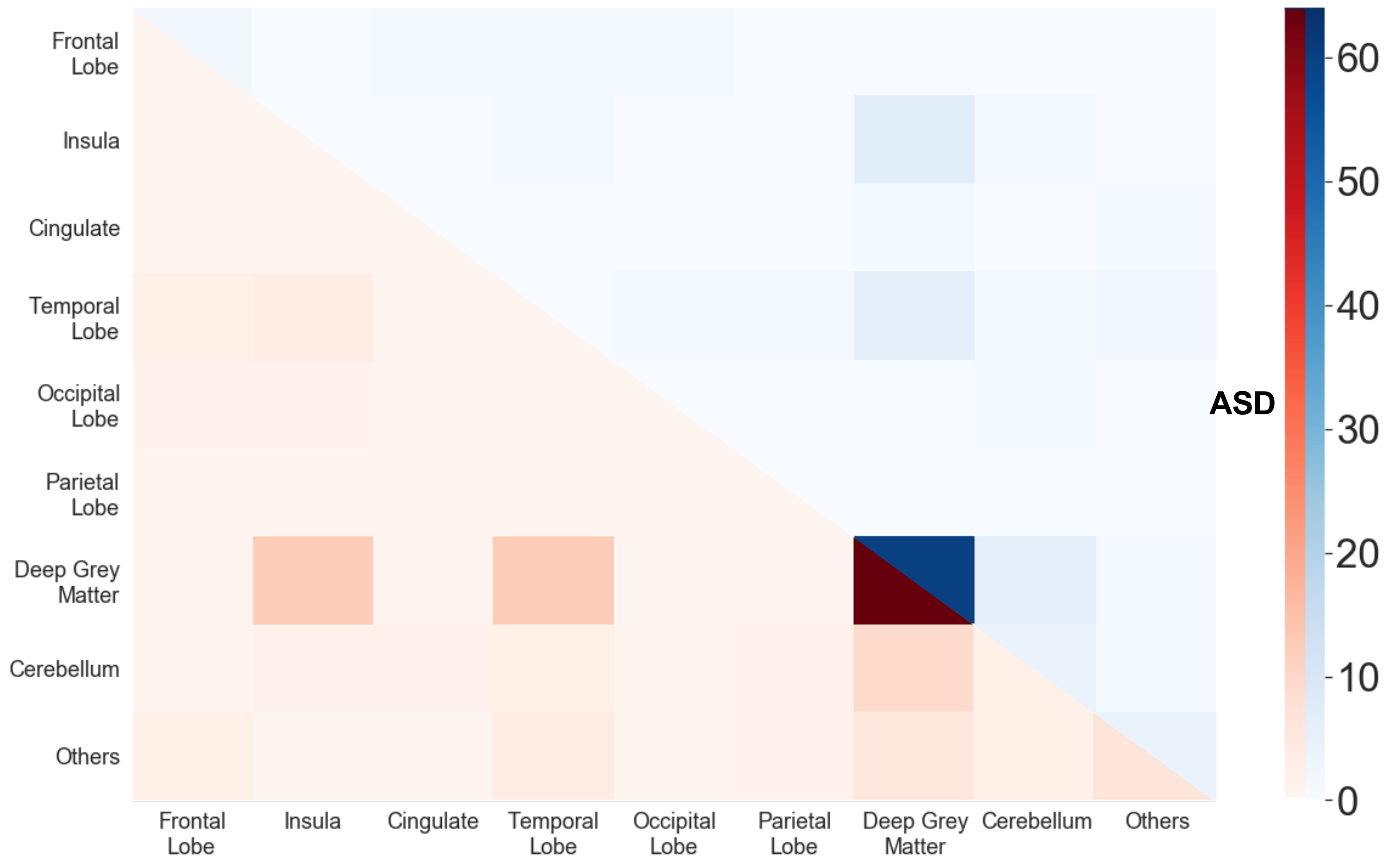} &    \includegraphics[width=0.32\textwidth]{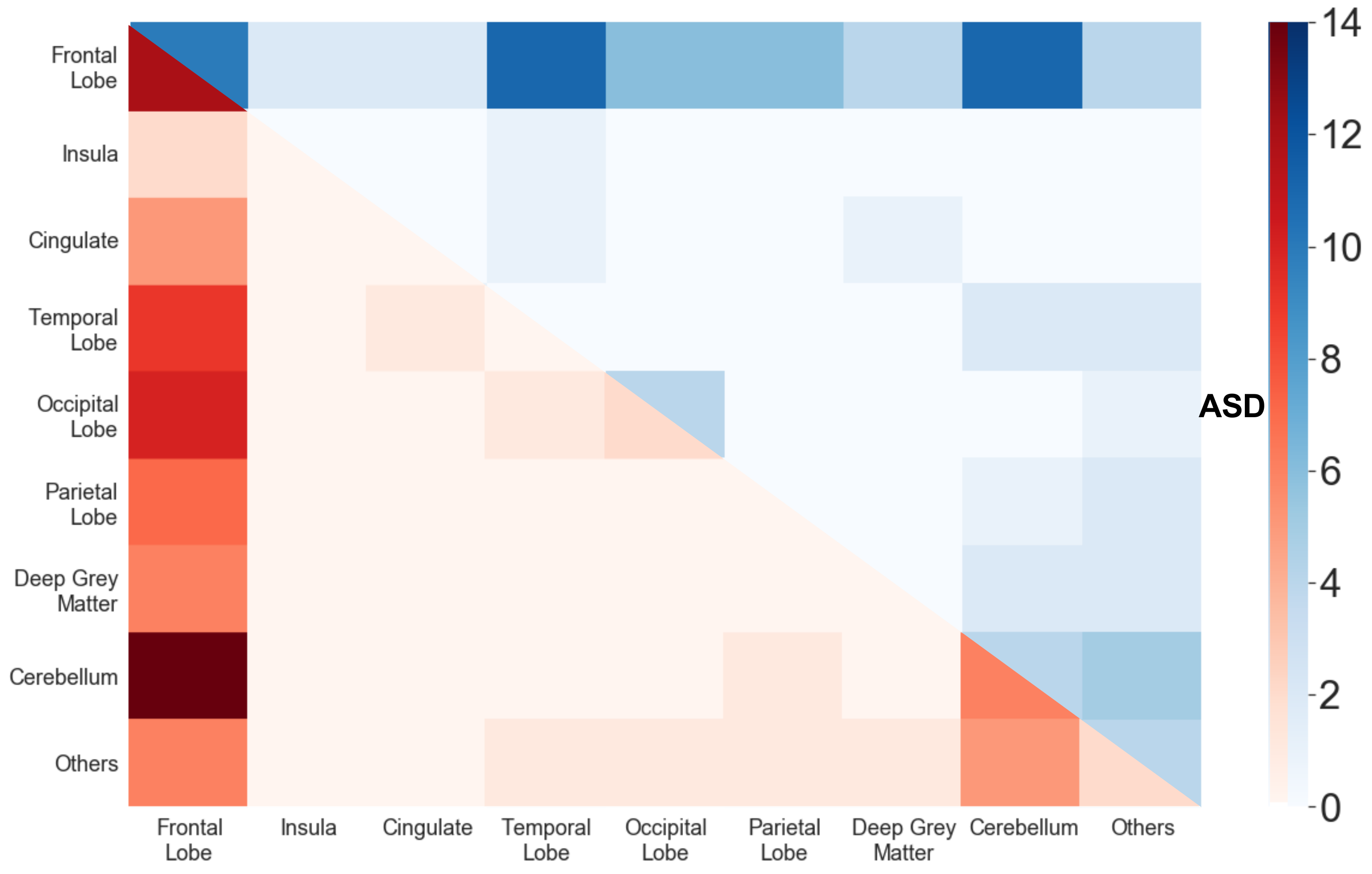} &
    \includegraphics[width=0.32\textwidth]{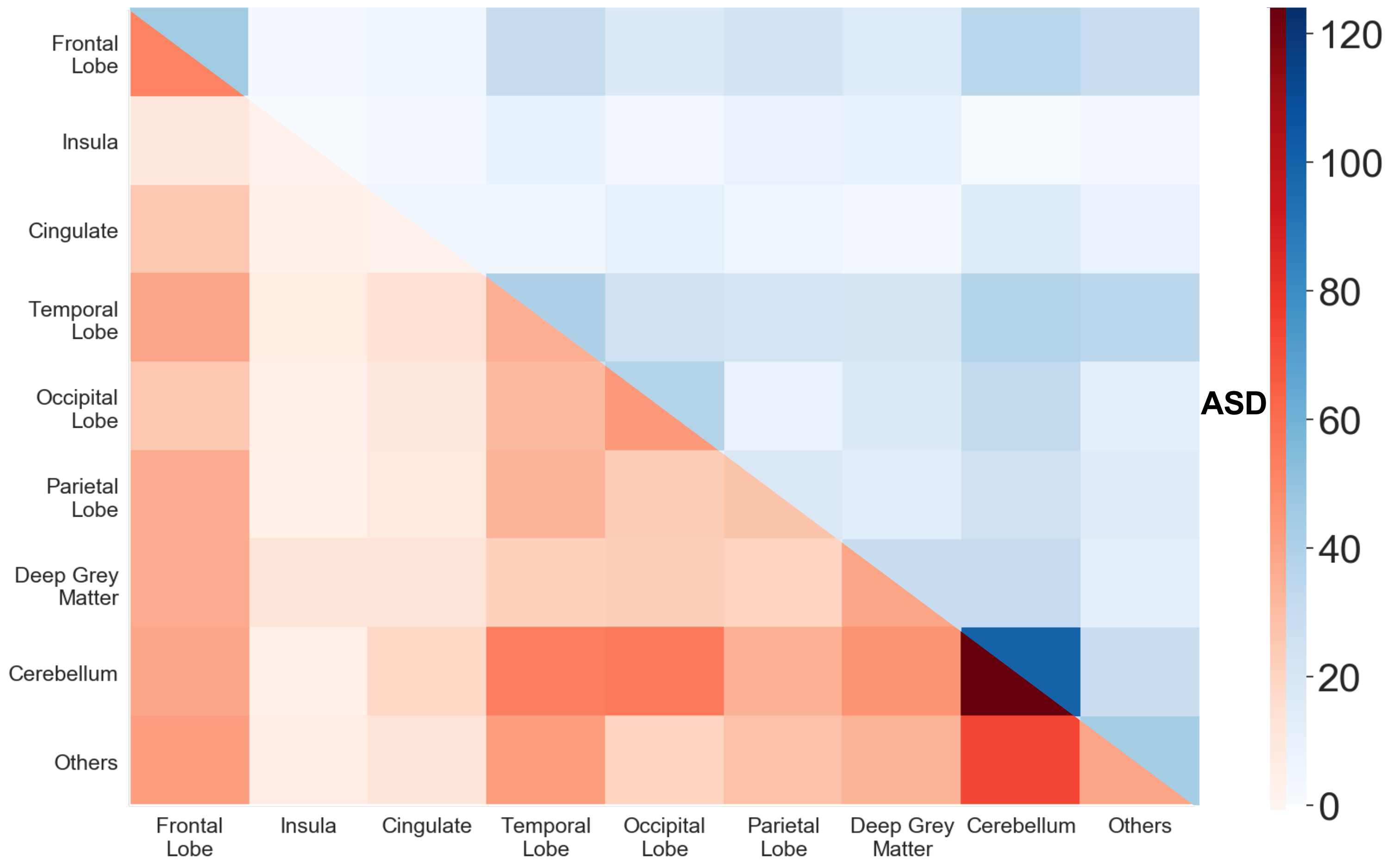}\\
   \includegraphics[width=0.32\textwidth]{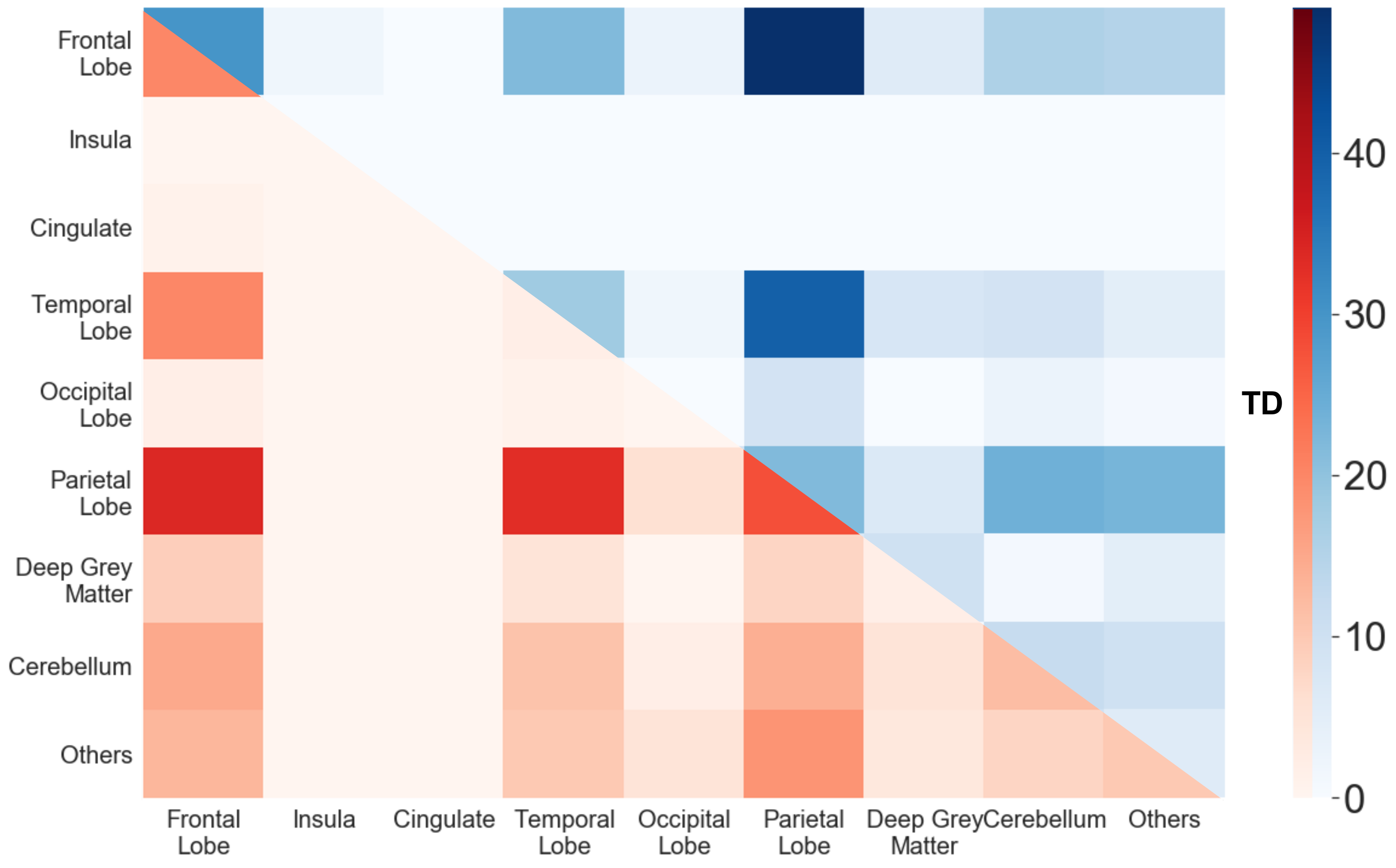} &
    \includegraphics[width=0.32\textwidth]{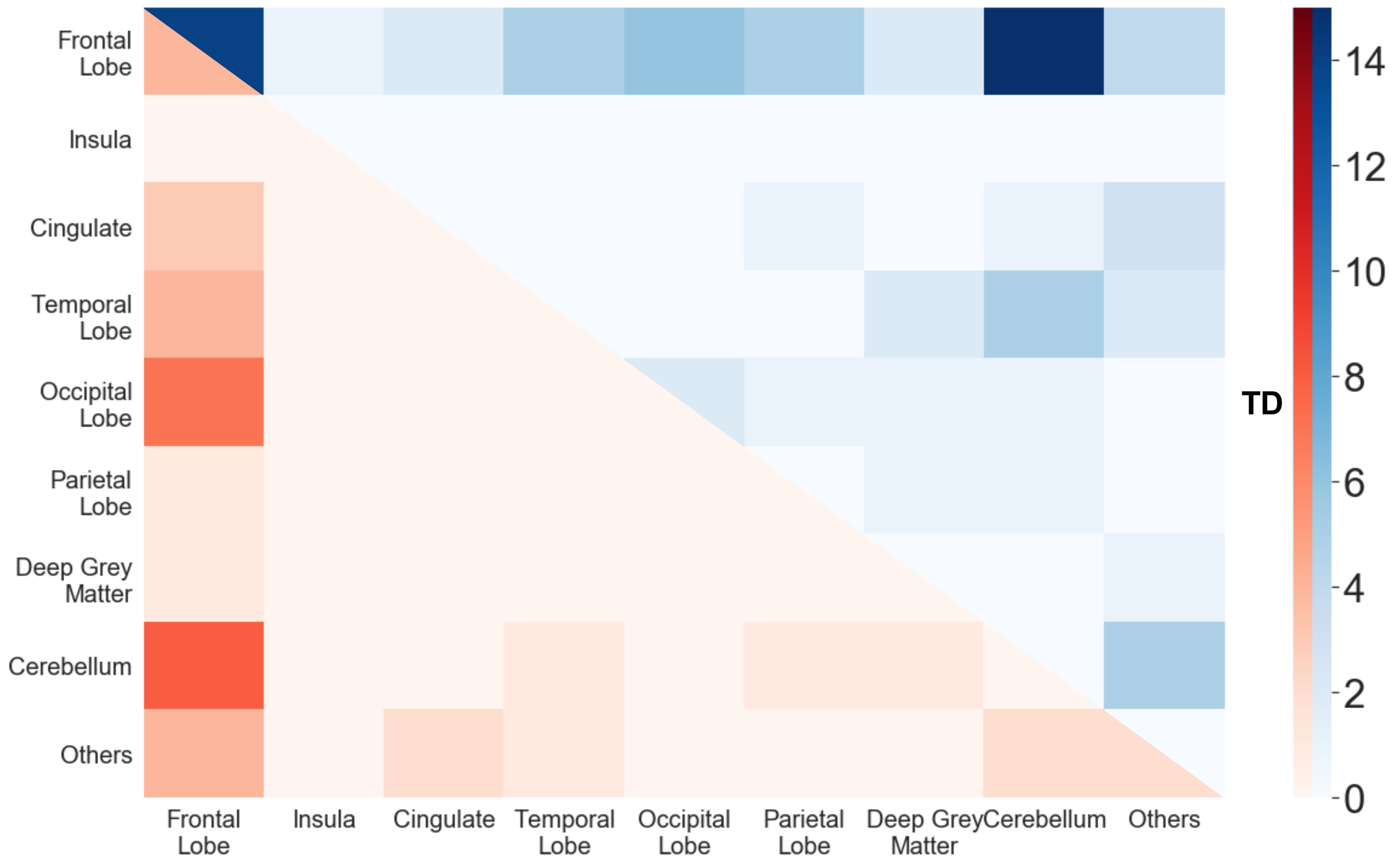} &
     \includegraphics[width=0.32\textwidth]{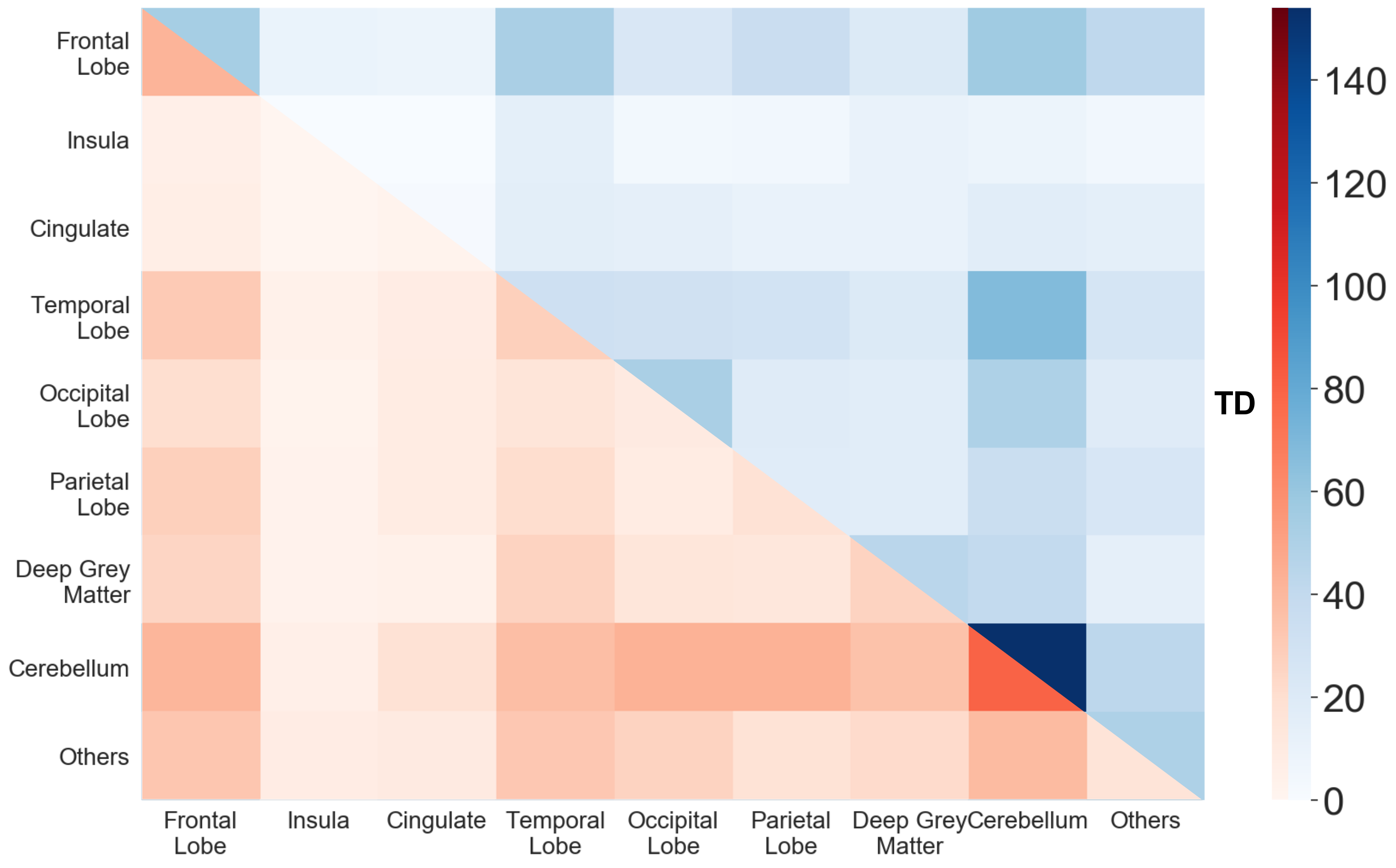}
    \end{tabular}
      \vspace{-4mm}
    \caption{Heatmaps representing global explanations for three different black-box classifiers on the \textsf{ASD} dataset, obtained by aggregating edges importance at the level of coarser-grain brain regions.}
    \label{fig:count_edges}
  \vspace{-2mm}
\end{figure*}
\subsection{Global explanations: edge-level aggregation} \label{subsec:appl:global1}
Similarly to what done above, we can aggregate the counterfactual graphs produced for many input graphs of both classes, so to produce statistics usable as a global explanation of the whole behaviour of a given black-box classifier.
Given a dataset of brain networks $D=\{E_{1},\dots,E_{n}\}$ all defined over the same set of {\roi}s $V$, similarly to what done previously in \S\ref{subsec:appl:local}, we can create one or more counterfactuals for each $E_i \in D$, then collect statistics about the occurrences of edges in the symmetric differences between the graphs and their counterfactuals. More formally, for each edge $e \in V^2$ we can define 4 counters:
\squishlist
\item $C_0^+(e) = | \{ (E,E_c) | f(E) = 0 \wedge e \in E^{c}_{+}\}|$, i.e., the number of pairs (original graph, counterfactual graph), such that the fact that $e \notin E$ seems to be important to explain $f(E)=0$;
\item $C_0^-(e) = | \{ (E,E_c) | f(E) = 0 \wedge e \in E^{c}_{-}\}|$; i.e., the number of pairs (original graph, counterfactual graph), such that the fact that $e \in E$ seems to be important to explain $f(E)=0$;
\squishend
and similarly $C_1^+(e)$ and $C_1^-(e)$  for class 1.
Once we have collected these statistics we can combine or plot them in several ways to provide meaningful global explanations. In Figure \ref{fig:count_edges} we present one such possible visualizations by means of heatmaps, aggregating {\roi}s in coarser-grain brain regions as provided by the AAL~\cite{aal} parcellation atlas, for three different black-box classifiers. In particular, assuming that \textsf{ASD} is class 1 and \textsf{TD} is class 0, in the heatmaps marked \textsf{ASD} (top) we report in the upper-triangle (blue) the aggregation of counters $C_0^+(e)$ and in the lower-triangle (red) the counters $C_1^-(e)$.
In other terms, both in the blue part that in the red part, the darker cells are those whose edges presence is important for the \textsf{ASD} class. For the way the 4 counters are defined, we expect to see a high level of symmetry in these heatmaps, as this is indeed the case in all 6 heatmaps in Figure \ref{fig:count_edges}.

A bird-eye view of the heatmaps highlights the fact that, \emph{although trained on the same training set with the same class labels, the three models have a very different inner logic, as unveiled by occurrences of the edges in the counterfactual explanations}.

In the case of the \textsf{Contrast Subgraphs} model, we can appreciate the relevance of \emph{deep grey matter } for the \textsf{ASD} class, consistently with the example in Figure \ref{fig:count_edges_local} and with neuroscience literature \cite{Richards_deep}. \emph{Paretal lobe}, \emph{temporal lobe} and \emph{frontal lobe} are instead the most important regions for the classification in the \textsf{TD} class. In fact, in the latter, the researchers found that the connectivity of \emph{Fusiform Gyrus}, that is in the temporal lobe, captures the risk of developing autism as early as 1 year of age and provides evidence that abnormal fusiform gyrus connectivity increases with age \cite{frontal_lobe,fusiform_gyrus}.
Also in the \textsf{Autoencoders} model, the \emph{frontal lobe} region is the most important to explain the classification. Finally, the heatmaps for the classifier based on \textsf{Graph2Vec} highlight the relevance of the \emph{cerebellum}, whose role in cognitive impairments in \textsf{ASD} has been observed in a plethora of studies, e.g., \cite{review2,cerebel_2,cerebel_3}.

\subsection{Global explanations: \roi-level aggregation} \label{subsec:appl:global2}
Finally, global explanations can also be produced in terms of importance of the {\roi}s for the classification, where the importance of a vertex can be computed by aggregating the same 4 counters defined in \S\ref{subsec:appl:global1} (i.e., $C_0^+$, $C_0^-$, $C_1^+$, and $C_1^-$), for all its incident edges.

\begin{figure}[t!]
    \centering
    \begin{tabular}{c}
    \textsf{Contrast Subgraphs}\\
    \includegraphics[width=1.0\linewidth]{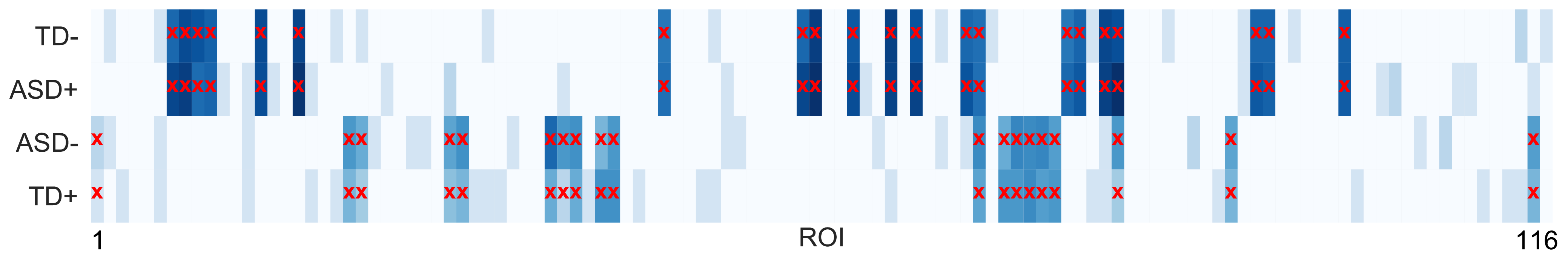}\\
    \textsf{Autoencoders}\\
    \includegraphics[width=1.0\linewidth]{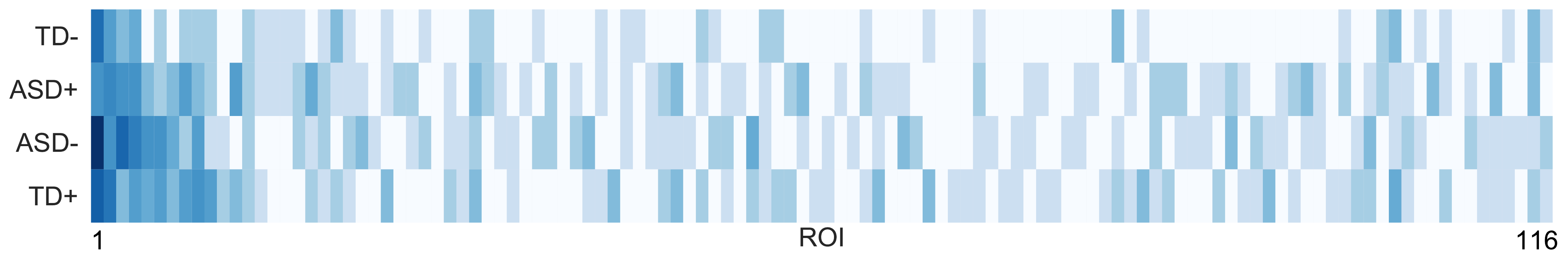}
    \end{tabular}
  \vspace{-4mm}
    \caption{Global explanation based on {\roi}s' importance:  \textsf{Contrast Subgraphs} black-box (top) and \textsf{Autoencoders}  black-box (bottom) on the \textsf{ASD} dataset (same setting as in Figure \ref{fig:count_edges}). On $x$-axis the 116 {\roi}s of the AAL atlas ordered by id, and on the $y$-axis the aggregation of the 4 counters defined in \S\ref{subsec:appl:global1} for all the edges incident on a given \roi. }
    \label{fig:auto_nodes_count}
\end{figure}

In Figure \ref{fig:auto_nodes_count} we provide two examples of explanations based on {\roi}s' importance.

\spara{The final ``litmus test''.} As we have a way to measure the importance of each single vertex in a black-box classifier, we can use the same idea we used in \S\ref{subsec:exp:whitebox}: i.e.,  to apply our counterfactual graphs framework to explain a ``white-box'' --  a classifier of which we know, as a sort of ground truth, the inner logic -- and check whether the most important vertices turn out to be the expected ones.

Therefore, we consider again the model based on \textsf{Contrast Subgraphs} on the \textsf{ASD} dataset. We recall that, by applying the method in \cite{lanciano2020}, we extract a contrast subgraph \textsf{ASD\_TD}, i.e., a set of vertices whose induced subgraph is very dense among \textsf{ASD} individuals and very sparse among \textsf{TD} individuals, and similarly, another
contrast subgraph \textsf{TD\_ASD}, i.e., a set of vertices whose induced subgraph is very dense among \textsf{TD} individuals and very sparse among \textsf{ASD} individuals. Then each graph in the dataset is mapped to two dimensions: the number of edges in the subgraph induced by each of the contrast subgraphs. Essentially the classification model is entirely defined by two sets of vertices and the classification hyperplane.

In Figure \ref{fig:auto_nodes_count} (top) we mark with red crosses the {\roi}s belonging to the  \textsf{TD\_ASD} contrast subgraph on the first two rows, and the {\roi}s forming the  \textsf{ASD\_TD} contrast subgraph on the third and fourth rows. The red crosses fall exactly on the darker cells of the heatmap: \emph{the global explanation built on counterfactual graphs is capturing exactly the (known and simple) logic of the classifier}.

\section{Conclusions and Future Work}
\label{sec:conclusions}
This paper introduces counterfactual graphs as a way to provide explanations of any black-box graph classifier, in the setting of graphs with node identity awareness, such is the case with brain networks. Counterfactual graphs can be used as basic building blocks to produce various type of explanations: they can be used directly as contrastive case-based explanations, they can be aggregated to provide LIME-like (SHAP-like) local explanations, as well as to produce global explanations.

We define the \textsc{Counterfactual Graph Search} problem and propose heuristic-search methods for it, distinguishing between two cases: when we do not have any data (oblivious), or when we have access to a dataset (data-driven). Our empirical assessment on different brain network datasets confirms that our methods produce good counterfactual graphs and that counterfactual graphs are a useful tool for explainability of brain network classification.

The language of explanations adopted in this paper is the most fine-grained possible: the edges whose addition or removal are important for a graph classification. In our future work, we plan to consider counterfactual graphs built over a richer vocabulary, including graph properties and higher-order structures, such as, e.g., motifs. On-going collaboration with neuroscientists is aimed at validating the usability of counterfactual graphs as a tool for explorative differential case-based reasoning for the domain expert.


\newpage
\appendix
\section{Reproducibility}
\label{sec:appendix}
Our Python code and the notebooks of our experiments are available at: \url{https://github.com/carlo-abrate/CounterfactualGraphs}.

The main pipeline of our experiments (Section \ref{sec:experiments}) is as follows:
\begin{description}
	\item[\textbf{Input:}] The input of the experiments is a dataset of labeled networks $D = \{(E_i,y_i)\}_{i=1}^n$ s.t. $E_i \in \mathbb{G}(V), y_i \in \{0,1\}$.
	\item[\textbf{Embedding training:}] We use multiple embedding techniques, whose main parameters are discussed in the rest of this Appendix.
	\item[\textbf{Classifier training and testing:}] The black-box oracle function $f$ is built by training a classifier on the embeddings. In particular, we tested support-vector machine (SVM), k-nearest neighbors (KNN), and Random Forest on each combination of embeddings parameters setting, using a 5-fold cross validation.
 Finally,  we selected the best  dimension of the embedding and the classification algorithm based on accuracy. In the paper we presented results based on KNN in combination with all the embedding methods, with the exception of \textsf{Contrast Subgraphs} where we use SVM.
\end{description}

As discussed in Section \ref{sec:algorithms}, counterfactual search methods iteratively call the oracle function for new candidate counterfactual graphs. The oracle function is composed of the embedding and the classifier part, at each call the candidate counterfactual graph is considered by the embedding as a new and unseen data. Some transductive methods do not naturally generalize to unseen data, due to non-determinism. However, for both \textsf{Graph2Vec} and \textsf{Sub2Vec}  we force the determinism as explained in\\ \url{https://github.com/RaRe-Technologies/gensim/issues/447}.

We next report the parameters settings used in the paper.

\spara{\textsf{Contrast Subgraphs} \cite{lanciano2020}.}  The code can be found at \url{https://github.com/tlancian/contrast-subgraph}.
We use the parameter $\alpha=0.025$ to generate the contrast subgraphs for \textsf{ASD} dataset and $\alpha=0.01$ for the \textsf{ADHD} dataset.
The accuracy is $0.77$ and $0.69$ for ABIDE and USCB datasets.

\spara{\textsf{Graph2Vec}\cite{annamalai2017graph2vec}.} The code can be found at \url{https://github.com/benedekrozemberczki/graph2vec}.
The main parameters used in the experiments of Table \ref{tab:agnosticmethod} for both \textsf{ASD} and \textsf{ADHD} datasets, are as follows:
\begin{itemize}
	\item $node\_degree =2$;
	\item $workers = 1$;
	\item $epochs = 100$;
    \item $min\_count = 5$;
    \item $learning\_rate = 0.025$;
    \item $down\_sampling = 0.0001$;
    \item vector space dim $= 50$ for both the datasets.
\end{itemize}
The accuracy is $0.66$ and $0.62$ for ABIDE and USCB datasets.

\vfill\eject 

\spara{\textsf{Sub2Vec} \cite{adhikari2018sub2vec}.} The code can be found at \url{ https://goo.gl/Ef4q8g}. The main parameters used in the experiments of Table \ref{tab:agnosticmethod} for both \textsf{ASD} and \textsf{ADHD} datasets, are as follows:
\begin{itemize}
	\item $type=Neighborhood$;
	\item $walk\_length = 100000$;
	\item $model = DBON$;
	\item $iteration=50$;
    \item vector space dim $= 50$ for both the datasets.
\end{itemize}
The accuracy is $0.65$ and $0.64$ for ABIDE and USCB datasets.

\textbf{Autoencoders \cite{gutierrez2019embedding}.}
The code can be found at \url{https://github.com/leoguti85/GraphEmbs}.
The main parameters used in the experiments of Table \ref{tab:agnosticmethod} for both \textsf{ASD} and \textsf{ADHD} datasets, are as follows:
\begin{itemize}
	\item $noise = 0.05$;
	\item $nb\_epoch = 150$;
	\item $batch\_size = 128$;
	\item $lr = 0.001$;
	\item vector space dim $= [200 ,400]$ for \textsf{ASD} and \textsf{ADHD} respectively.
\end{itemize}
The accuracy is $0.68$ and $0.63$ for ABIDE and USCB datasets.

We finally summarize report the main parameters and settings used to generate the figures of the paper:
\begin{description}
	\item[Fig. \ref{fig:intro}] Dataset: \textsf{ASD}; Patient: \texttt{USM\_0050453}; Classifier: \textsf{Contrast Subgraphs} with SVM; Counterfactual Search: \textsf{OFS+OBS} with $k=5$, $\eta=2000$.
	\item[Fig. \ref{fig:methods1} and \ref{fig:methods2}] Dataset: \textsf{ASD}; Classifier: \textsf{Contrast Subgraphs} with SVM; Counterfactual Search: \textsf{DS}, \textsf{OFS+OBS} and \textsf{DFS+DBS}, with $k=5$, $\eta=2000$.
	\item[Fig. \ref{fig:optimumcount}] Dataset: \textsf{ASD}; Classifier: \textsf{Contrast Subgraphs} with SVM; Counterfactual Search\textsf{ OFS+OBS}, with $k=5$, $\eta=2000$. Patients in the zoomed section are: \texttt{UM\_1\_0050366}, \texttt{Stanford\_0051164},\texttt{UCLA\_1\_0051226}.
	\item[Fig. \ref{fig:count_edges_local}] Dataset: \textsf{ASD}; Patient: \texttt{MaxMun\_d\_0051353}; Classifier: \textsf{Contrast Subgraphs} with SVM; Counterfactual Search:\textsf{ OFS+OBS} with $k=5$, $\eta=2000$.
	\item[Fig. \ref{fig:count_edges}, \ref{fig:auto_nodes_count}] Dataset: \textsf{ASD}; Classifier: \textsf{Contrast Subgraphs} with SVM, \textsf{Autoencoders, Graph2Vec} with KNN; Counterfactual Search: \textsf{OFS+OBS}, with $k=5$, $\eta=2000$.
\end{description}

\section{Acknowledgements}
Authors acknowledge support from Intesa Sanpaolo Innovation Center.
The funders had no role in study design, data collection and analysis, decision to publish, or preparation of the manuscript.

\end{document}